\documentclass[aps,pre,superscriptaddress, amsmath, amssymb, showpacs, twocolumn]{revtex4-1} 

\newcommand{\B}{\textrm{B}}
\newcommand{\massrat}{\mathrm{m_{r}}}

\newcommand{\vc}{\mathbf}
\usepackage{graphicx}
\usepackage{color}
\begin{document}

\title{Effects of Coulomb Coupling on Stopping Power and a Link to Macroscopic Transport}

\author{David J. Bernstein}
\author{Scott D. Baalrud}
\affiliation{Department of Physics and Astronomy, University of Iowa, Iowa City, Iowa 52242, USA}
\author{J\'er\^ome Daligault}
\affiliation{Los Alamos National Laboratory, Los Alamos, New Mexico 87545, USA}

\date{\today}

\begin{abstract}
Molecular dynamics simulations are used to assess the influence of Coulomb coupling on the energy evolution of charged projectiles in the classical one-component plasma. 
The average projectile kinetic energy is found to decrease linearly with time when $\nu_{\alpha}/\omega_{p} \lesssim 10^{-2}$, where $\nu_{\alpha }$ is the Coulomb collision frequency between the projectile and the medium, and $\omega_{p}$ is the plasma frequency. 
Stopping power is obtained from the slope of this curve. 
In comparison to the weakly coupled limit, strong Coulomb coupling causes the magnitude of the stopping power to increase, the Bragg peak to shift to several times the plasma thermal speed, and for the stopping power curve to broaden substantially.
The rate of change of the total projectile kinetic energy averaged over many independent simulations is shown to consist of two measurable components: a component associated with a one-dimensional friction force, and a thermal energy exchange rate. 
In the limit of a slow and massive projectile, these can be related to the macroscopic transport rates of self-diffusion and temperature relaxation in the plasma. 
Simulation results are compared with available theoretical models for stopping power, self-diffusion coefficients, and temperature relaxation rates.  

\end{abstract}

\maketitle
\section{Introduction}
\label{sec:intro}

This work assesses fundamental concepts associated with the use of stopping power as a description of the energy loss rate of charged projectiles in strongly coupled plasmas. 
Despite many years of investigation, stopping power remains an interesting topic with many open questions.
In particular, while stopping power is well characterized in classical weakly coupled plasmas, recent work is concerned with dense and strongly coupled plasmas (e.g. ~\cite{Fraley1974,Mehlhorn1981,Petrasso,Zwick_Nuovo,Zwick_Strong1,Zwick_Review,MD_Grabowski}), such as those in inertial confinement fusion (ICF) experiments. 
Conventional theories do not apply to dense plasmas because electrons are not classical and ions are not weakly coupled.
The problem is interesting not only because of its inherent importance to fusion energy research, but also because stopping power can offer a unique diagnosis of complicated processes occurring in dense plasmas that are not easily accessible from more commonly-studied equilibrium properties.
For that matter, it has been proposed that stopping power can provide an indirect way to measure macroscopic transport properties, such as diffusion \cite{Dufty1, Dufty2} or temperature relaxation, which are often difficult to measure directly. 
This work describes the influence of Coulomb coupling on stopping power, as well as methods to relate stopping power to macroscopic transport rates.

To this end, the energy evolution of a charged projectile evolving in the classical one-component plasma (OCP) is studied using molecular dynamics (MD) simulations. 
The OCP consists of a system of identical electrically charged point particles interacting through the Coulomb potential and immersed in a rigid, uniform background of opposite charge to ensure overall charge neutrality~\cite{BausHansen}. 
Although the OCP does not include all of the relevant aspects of a dense plasma, such as multiple species and electron physics, it isolates the effects of strong Coulomb coupling. 
This allows us to test theoretical models that have been proposed to extend traditional stopping power theory to strongly coupled plasmas.  
Coulomb coupling in the OCP is characterized by the ratio of the Coulomb potential energy at the average interparticle spacing to the average kinetic energy 
\begin{equation}
\Gamma \equiv \frac{q^2/a}{k_BT}. \label{gamma}
\end{equation}
The plasma is considered strongly coupled if $\Gamma \gtrsim 1$. 
Here, $q$ is the electrical charge, $a \equiv (4\pi n/3)^{-1/3}$ is the average inter-particle spacing, $n$ is the plasma number density, $k_{\B}$ is the Boltzmann constant, and $T$ is the plasma temperature.
In thermal equilibrium at temperature $T$, the OCP is fully characterized by the coupling parameter $\Gamma$.

Using MD simulations, the applicability of stopping power as a means to characterize the energy evolution of projectiles is assessed for a broad range of projectile speeds, masses and coupling strength of the background plasma.
The projectiles are assumed to have a single charge of the same sign as a background plasma particle's. 
The energy evolution can be described by the stopping power when the projectile energy changes linearly in time shortly after being introduced into the plasma. 
Such behavior is found to require that the projectile-plasma Coulomb collision rate be sufficiently slow compared to the plasma frequency ($\nu_{\alpha }/\omega_p \lesssim 10^{-2}$, where $\omega_p = \sqrt{4\pi q^2 n/m}$). 
At strong coupling, it is also found to require that the initial projectile kinetic energy be greater than the potential energy of the background plasma ($\frac{1}{2} m_\alpha V_0^2 \gtrsim q^2/a$, where $m_\alpha$ and $V_0$ are the projectile mass and speed, respectively). 
When these conditions are met, stopping power is computed from the simulations.
Strong coupling is found to increase the magnitude of the stopping power, cause the Bragg peak to shift to higher speed and cause the stopping power curve to broaden substantially. 

Simulation results are compared with two types of theoretical models: Those based on linear response theory and those based on discrete Coulomb collisions. 
Models for extending these traditional approaches to strong coupling include various approximations for the linear response function using a local field correction in the dielectric models~\cite{Zwick_Review,Ichimaru}, and the effective potential theory (EPT) for the collision model~\cite{EPTRL,EPT}. 
All models are found to accurately predict the MD data at weak coupling ($\Gamma = 0.1$), but the linear response models are more accurate for fast projectile speeds. 
This is attributed to the influence of dynamic screening. 
At strong coupling, the EPT model is generally found to be more accurate regardless of the projectile speed. 
In particular, the models based on linear dielectric response are found to rapidly break down at strong coupling, predicting a very low value of stopping power for slow projectiles that do not agree with the MD simulation results. 

Stopping power is also analyzed from a new viewpoint by splitting the total energy into a component associated with the frictional slowing of the projectile along its initial direction, and a second component associated with the energy gained in coming to thermal equilibrium with the background. 
Whereas the traditional viewpoint of stopping power considers only the former of these components, it is shown that the latter component can be significant for sufficiently slow projectile speeds or when the mass ratio between the projectile and background plasma species ($\massrat =m_\alpha/m$) is near unity. 
This interpretation resolves a discrepancy between previous models based on Coulomb collisions~\cite{Binary_FA} and those based on linear-response theory.

A connection between the low-speed region of the stopping power curve for a massive projectile and macroscopic transport rates is also explored. 
In particular, the EPT model is used to predict a connection between the frictional component of the stopping power for a massive projectile and the self-diffusion coefficient of the background plasma (or the mutual diffusion coefficient for a massive impurity species). 
Similarly, it is also used to predict a connection between the thermal energy component and the temperature relaxation rate of the background plasma. 
These predictions are found to be in good agreement with the MD simulation data, as well as MD simulation results from the literature where these rates are computed using independent methods. 
This connection may provide additional means to infer macroscopic transport rates from a measurement of the microphysical process of individual projectile stopping. 


\section{Theory}
\label{sec:models}

\subsection{Conventional notion of stopping power\label{sec:conv}} 

Stopping power is defined as the negative spatial rate of change of the average kinetic energy of a projectile as it traverses a medium: $-d\mathcal{E}/dx$.
It is typically expressed as a function of the projectile speed $V_0$, and is related to the instantaneous time rate of change of the projectile kinetic energy by~\cite{Zwick_Review}
\begin{equation}
\label{eq:dEdx_dEdt}
-\frac{d\mathcal{E}}{dx} = -\frac{1}{V_0}\frac{d\mathcal{E}}{dt}.
\end{equation}
This concept is usually applied to the situation where the projectile can be modeled as slowing in the single dimension co-aligned with its initial velocity, in which case it is the friction force opposing the motion of the projectile. 
We will denote this 1D friction force $-d\mathcal{E}_\textrm{F}/dx$ to distinguish it from an alternative formulation described in Sec.~\ref{sec:tc}. 

In a plasma, stopping power can be modeled by using the linear dielectric model. 
In this model, the charged projectile generates a wake in a background plasma modeled as a linear dielectric medium, and the force on the projectile is computed from the induced electric field associated with this wake~\cite{Ichimaru} 
\begin{equation}
 \label{eq:dEdx_dielectric}
\frac{d\mathcal{E}_F}{dx} \biggl|_{\textrm{D}} = \frac{q^{2}}{\pi V_{0}^{2}} \int_0^{k_{\textrm{max}}} \frac{dk}{k}\int_{-kV_{0}}^{kV_{0}} d\omega\, \omega \operatorname{Im}\Bigg\{ \frac{1}{\varepsilon(k,\omega)} \Bigg\}.
\end{equation}
Here, $k$ is the magnitude of the wave vector, $q$ is the charge of the projectile (which the same as a background particle's), and $\varepsilon(k,\omega)$ is the linear dielectric response function of the plasma.
Because the dielectric model does not explicitly treat particle collisions, the integral over $k$ diverges for high wave numbers (short distances).
A cut-off is typically chosen as the inverse distance of closest approach between the projectile and background particles, $k_{\max} \approx \mu (v_{T}^{2}+V_0^{2})/2q^{2}$, where $\mu \equiv m_\alpha m/(m_\alpha+m)$ is the reduced mass and $v_T = \sqrt{2k_\B T/m}$ is the thermal speed of the background~\cite{Zwick_Review}.

If the plasma is weakly coupled ($\Gamma \ll 1$) and near thermal equilibrium, the random-phase-approximation (RPA) provides an accurate description for the dielectric function~\cite{Ichimaru}
\begin{equation}
\label{eq:weakeps}
\varepsilon_{\textrm{RPA}}(k, \omega) = 1 - U(k)\chi _{0}(k,\omega).
\end{equation}
Here, $U(k) \equiv 4\pi q^{2}/k^{2}$ is the Fourier transform of the bare Coulomb potential, $\chi_{0}(k, \omega) \equiv  \frac{n}{2k_{B}T}Z'(\frac{\omega}{kv_{T}})$ is the linear response function, and $Z'(x) = dZ(x)/dx$ is the derivative of the plasma dispersion function [$Z(x)$]~\cite{frie:61}. 

A number of models have been proposed to extend this to strong coupling.   
These are typically expressed in the form~\cite{Ichimaru} 
\begin{equation}
\label{eq:LFC_dielectric}
\varepsilon(k,\omega) = 1-\frac{U(k)\chi_{0}(k,\omega)}{1+U(k) G(k,\omega)\chi_{0}(k,\omega)},
\end{equation}
where $G(k, \omega)$ is the local field correction (LFC). 
The burden in these models is to find an accurate expression for the LFC. 

In the low-frequency limit, the LFC is directly related to the static structure factor of the plasma $[ S(k) ]$ ~\cite{Ichimaru}
\begin{equation}
\label{eq:LFC}
G_\textrm{S}(k) \equiv G(k,\omega \rightarrow 0) = 1 - \frac{a^{2}k^{2}}{3\Gamma}\Big[\frac{1}{S(k)}-1\Big],
\end{equation}
where $S(k) = 1+nh(k)$, and $h(k)$ is the pair-correlation function~\cite{HansenMcDonald}.
In real space, $h(r) = g(r) - 1$, where $g(r)$ is the pair distribution function and $r$ the distance.
Good approximations have been developed for $g(r)$ in the OCP at essentially any coupling strength. 
One example is the hypernetted-chain (HNC) approximation~\cite{HansenMcDonald}
\begin{eqnarray}
\left\{
\begin{array}{l}
h(k) = c(k)[1+nh(k)] \\
g(r) = \exp[-U(r)/k_{B}T+h(r)-c(r)] ,
\end{array}
\right.
\label{eq:OZ_HNC}
\end{eqnarray}
where $c(r)$ is the direct correlation function. 

Approximations for the LFC beyond the static approximation are very challenging. 
For this reason, the LFC is usually modeled by simply using the static limit $[G_\textrm{S}(k)]$ in Eq.~(\ref{eq:LFC_dielectric}).  
However, approximations for dynamic LFCs have been proposed~\cite{Ich_DLFC,HK_DLFC}, which attempt to match exact asymptotic results from the low and high frequency limits. 
A comparison of stopping power computed from each of these models is provided in Appendix~\ref{sec:DLFC}. 
This shows that differences in the prediction of each model are modest, and influence only high speed projectiles. 
Previous tests of the LFC models using MD simulations have shown that neither provides an accurate approximation covering the broad coupling, frequency and wavenumber range of interest in this work~\cite{Mithen}. 
For this reason, it is not known which, if either, model provides a more reliable prediction. 
Aside from Appendix~\ref{sec:DLFC}, all results shown for the dielectric model will be based on the static LFC [$G_\textrm{S}(k)$].

\subsection{Stopping power as a two-component process\label{sec:tc}}

Both the conventional notion of stopping power and the dielectric model consider one-dimensional slowing of projectiles along their initial velocity vector ($\vc{V}_0$). 
However, in reality collisions cause a projectile to acquire kinetic energy perpendicular to $\vc{V}_0$ in addition to slowing down. 
If one considers a large statistical sample of independent projectiles with identical initial conditions evolving in identical plasmas (with different particle configurations), the variance of the kinetic energy of the sample about the mean provides additional information about the projectile energy evolution. 

To quantify this, consider that the average kinetic energy of a sample of projectiles can be split into two components
\begin{equation}
\label{eq:energy}
\mathcal{E} \equiv \frac{1}{2}m_\alpha \langle \vc{v}^2 \rangle = \mathcal{E}_\textrm{F} + \mathcal{E}_\textrm{T} .
\end{equation}
where $\mathcal{E}_\textrm{F} \equiv \frac{1}{2}m_\alpha V^{2}$ is associated with the mean velocity (one-dimensional slowing), and $\mathcal{E}_\textrm{T} \equiv \frac{1}{2} m_\alpha \langle (\vc{v}-\vc{V})^2 \rangle$ is associated with the variance of the projectile speeds about the mean. 
Here $\vc{v}$ is the individual projectile velocity, and $\vc{V} = \langle \vc{v} \rangle$ is the mean velocity of the sample of projectiles. 
The averages in Eq.~(\ref{eq:energy}) can be expressed as
\begin{equation}
\label{eq:thermal_avg}
\langle H(\vc{v}) \rangle = \int d^{3}v H(\vc{v}) \mathcal{F}_{\alpha}(\vc{v},t),
\end{equation}
where the velocity distribution of projectiles is
\begin{equation}
\label{eq:proj_density}
\mathcal{F}_\alpha(\vc{v},t) = \frac{1}{\mathcal{N}}\sum_{i=1}^{\mathcal{N}} \delta[\vc{v}-\vc{v}_i(t)]. 
\end{equation}
Here, $\mathcal{N}$ is the total number of projectiles, and $\vc{v}_{i}(t)$ is the velocity of the $i^{th}$ projectile. 
We emphasize that the distribution of projectiles here represents the statistical average of independent ``trials''; only one projectile exists in the plasma at any given time. 
The concept may be extended to a plasma with many projectiles, as long as the projectile concentration is sufficiently dilute that they effectively do not interact with one another. 
Thus, the friction and thermalization terms associated with projectile slowing can also be associated with the similar processes of a dilute component of a mixture. 
Accounting for each category of the total kinetic energy, stopping power can be expressed as consisting of two components 
\begin{equation}
\label{eq:dEdx_total}
-\frac{d\mathcal{E}}{dx} = -  \frac{d\mathcal{E}_\textrm{F}}{dx} - \frac{d\mathcal{E}_\textrm{T}}{dx} . 
\end{equation}
The first corresponds to the conventional notion of stopping power from Sec.~\ref{sec:conv}, and the second is an additional component associated with thermalization of the projectile. 

To connect with kinetic theory, it is useful to express the time evolution of the projectile distribution as a kinetic equation, $d \mathcal{F}_{\alpha,0}/d t = C_{\alpha}(\vc{v})$, where $C_{\alpha}$ is the collision operator. 
Because the projectiles are initiated with the same velocity, $\mathcal{F}_{\alpha,0}(\vc{v}) = \delta(\vc{v}-\vc{V}_{0})$.
At early times, it follows that 
\begin{equation}
\label{eq:dEdt_energy}
\frac{d\mathcal{E}}{dt} = \vc{V}_0 \cdot \vc{R}_{\alpha}+Q_{\alpha},
\end{equation}
where
\begin{equation}
\label{eq:R}
\vc{R}_{\alpha} \equiv m_\alpha \int d^3v\, \vc{v} C_{\alpha}(\vc{v})
\end{equation} 
and
\begin{equation}
\label{eq:Qs}
Q_{\alpha} \equiv \frac{1}{2}m_\alpha \int d^3v\, (\vc{v}-\vc{V}_0)^2 C_{\alpha} (\vc{v}) .
\end{equation}
Using Eq.~(\ref{eq:dEdx_dEdt}), the stopping power can then be expressed in the form of Eq.~(\ref{eq:dEdx_total}), where
\begin{equation}
\label{eq:Stop_fric}
\frac{d\mathcal{E}_\textrm{F}}{dx} \equiv \frac{\vc{V}_0}{V_0} \cdot \vc{R}_{\alpha }.
\end{equation}
represents a one-dimensional friction force associated with slowing of the mean projectile speed, and 
\begin{equation}
\label{eq:Stop_therm}
\frac{d\mathcal{E}_\textrm{T}}{dx} \equiv \frac{Q_{\alpha }}{V_0}.
\end{equation}
represents a ``thermalization force'' associated with the rate at which the velocity of projectiles spread about the mean.  
The friction term will always be negative ($d\mathcal{E}_\textrm{F}/dx<0$) indicating that the mean projectile velocity is decreasing, whereas the thermalization term will always be positive ($d\mathcal{E}_\textrm{T}/dx>0$), representing that the variance of speeds about the mean increases in time. 
Models for each of these contributions based on different collision operators are described below. 

\subsection{Lenard-Balescu model}

A common plasma kinetic theory is provided by the Lenard-Balescu collision operator~\cite{Lenard,Balescu}
\begin{align}
\label{eq:CLB}
C_{\textrm{LB}} &= -\frac{2 q^4}{m_\alpha} \frac{\partial}{\partial \vc{v}} \cdot \int d^{3}k d^3v^\prime\, \frac{\delta [\vc{k} \cdot (\vc{v}-\vc{v'})]}{|\varepsilon_{\textrm{RPA}}(k,\vc{k}\cdot \vc{v})|^2} \\ \nonumber
& \frac{\vc{k}\vc{k}}{k^4} \cdot \Bigg[\frac{\mathcal{F}_{\alpha,0}(\vc{v})}{m}\frac{\partial f(\vc{v'})}{\partial \vc{v'}}-\frac{f(\vc{v'})}{m_\alpha}\frac{\partial \mathcal{F}_{\alpha,0}(\vc{v})}{\partial \vc{v}} \Bigg],
\end{align}
where $f$ is the background plasma distribution, which is assumed to be Maxwellian here. 
Applying the Lenard-Balescu collision operator to Eq.~(\ref{eq:R}), the frictional stopping power term is given exactly by Eq.~(\ref{eq:dEdx_dielectric}) where the dielectric function is the RPA dielectric function [Eq.~(\ref{eq:weakeps})].
This is expected since the Lenard-Balescu collision operator is derived from the same linear-response theory basis as was used to derive the friction force in Sec.~\ref{sec:conv}. 

In addition, using the collision operator Eq.~(\ref{eq:CLB}) in Eq.~(\ref{eq:Qs}) provides
\begin{equation}
\label{eq:dEdx_dieltherm}
\frac{d\mathcal{E}_\textrm{T}}{dx} \biggl|_{\textrm{LB}} = \frac{8\sqrt{\pi}q^4n}{m_\alpha v_{T}}\frac{e^{-V_0^2/v_{T}^2}}{V_0} \int_0^{k_{\textrm{max}}} \frac{dk/k}{|\varepsilon_{\textrm{RPA}}(k,kV_0)|^2}.
 \end{equation}
Although models for extending the plasma dielectric function to strong coupling have been provided, as described in Sec.~\ref{sec:conv}, the Lenard-Balescu equation has only been generalized to treat strong coupling in situations where the interacting distribution functions are near thermal equilibrium~\cite{Ichimaru}. 
Since this is not the situation for the interaction of a single projectile with a background plasma, we only apply the Lenard-Balescu result with the RPA dielectric function. 

\subsection{Collision model}

The second model we compare with is a collision model based on EPT~\cite{EPTRL,EPT}. 
This is a kinetic theory based on a Boltzmann-like collision operator
\begin{equation}
\label{eq:CB}
C_\textrm{B} = \int d^3v' d\Omega \sigma u [\mathcal{F}_{\alpha,0}(\hat{\vc{v}})f(\hat{\vc{v}}')-\mathcal{F}_{\alpha,0}(\vc{v})f(\vc{v'})],
\end{equation} 
but with a modified collision cross section that models the influence of strong coupling by treating binary interactions as occurring via the potential of mean force. 
Here, $d\Omega = d\phi d\theta \sin \theta$ is the solid angle,  $\sigma$ is the differential scattering cross section, $u = |\vc{v} - \vc{v}^\prime|$ is the relative speed, $\hat{\vc{v}}$ and $\hat{\vc{v}}'$ are the post-collision velocities, and $\vc{v}$ and $\vc{v}'$ are the pre-collision velocities. 
Applying Eq.~(\ref{eq:CB}) in Eq.~(\ref{eq:R}), the frictional stopping power term is~\cite{ConfProceedings}
\begin{equation}
\label{eq:Rept}
\frac{d\mathcal{E}_\textrm{F}}{dx}\biggl|_{\textrm{EPT}} = -m_{\alpha} V_0 \nu_{\alpha}
\end{equation}
where
\begin{equation}
\label{eq:collfreq}
\nu_{\alpha} = \frac{16\sqrt{\pi}n q^{4}}{3m_\alpha \mu v_{T}^{3}} \Xi^{[\textrm{F}]}_{\alpha} ( \bar{V}_0)
\end{equation}
is a collision frequency between the projectile and background particles, $\bar{V}_0 \equiv V_0/v_{T}$, and 
\begin{multline}
\label{eq:GenCoul}
\Xi^{[\textrm{F}]}_{\alpha}(\bar{V}_0) = \frac{3}{32} \frac{1}{\bar{V}_0^3} \int_0^\infty d\xi \xi^2 \frac{\bar{\sigma}(\xi)}{\sigma_o} \times \\ \times \biggl[ (2\xi \bar{V}_0+1) e^{-(\xi+\bar{V}_0)^2} + (2\xi \bar{V}_0 -1) e^{-(\xi-\bar{V}_0)^2} \biggr]
\end{multline}
is a velocity-dependent generalized Coulomb logarithm. 
Here $\xi \equiv |\vc{v}'-\vc{V}_0|/v_{T}$, $\sigma_o \equiv \pi q^4/\mu^2 v_{T}^4$, and
\begin{equation}
\label{eq:momscat}
\bar{\sigma}(v_{r}) = 2 \pi \int_{0}^{\infty} db b [1-\cos(\pi-2\Theta(b,v_{r}))]
\end{equation}
is the momentum-transfer scattering cross section, where $v_r = \xi v_{T}$, and $b$ is the impact parameter. 
The scattering angle is
\begin{equation}
\label{eq:scatangle}
\Theta(b,v_{r}) = b \int_{r_{0}}^{\infty} dr \frac{1}{r^{2}}\Bigg[1-\frac{b^{2}}{r^{2}}-\frac{2\phi(r)}{\mu v_{r}^{2}}\Bigg]^{-1/2} ,
\end{equation}
where $\phi(r)$ is the interaction potential, and $r_{0}$ is the distance of closest approach, which is determined by the largest root of the denominator of the integrand.

When Eq.~(\ref{eq:CB}) is applied to Eq.~(\ref{eq:Qs}), the thermalization term is
\begin{equation}
\label{eq:Qept}
\frac{d\mathcal{E}_\textrm{T}}{dx} \biggl|_{\textrm{EPT}} = \frac{8\sqrt{\pi}q^4n}{m_{\alpha}v_{T}}\frac{\Xi^{[\textrm{T}]}_{\alpha}(\bar{V}_0)}{V_0}, 
\end{equation}
where
\begin{equation}
\label{eq:GenCoul_2}
\Xi^{[\textrm{T}]}_{\alpha}(\bar{V}_0) = \frac{1}{8 \bar{V}_0} \int_0^\infty d\xi\, \xi^4 \frac{\bar{\sigma}(\xi)}{\sigma_o} \biggl[ e^{-(\xi - \bar{V}_0)^2} - e^{-(\xi + \bar{V}_0)^2} \biggr] 
\end{equation} 
is a different velocity-dependent generalized Coulomb logarithm~\cite{ConfProceedings}. 

In the weakly-coupled limit, $\bar{\sigma}/\sigma_{o} = 4\xi^{-4} \ln\Lambda$, where $\Lambda = 1/(\sqrt{3}\Gamma^{3/2})$ is the plasma parameter \cite{Binary_FA, EPT}.  
In this limit, the results presented above in Eqs.~(\ref{eq:Rept}) and (\ref{eq:Qept}) are identical to those derived by de~Farrariis and Arista~\cite{Binary_FA}. 
The EPT generalizes this result by extending it into the regime of strong Coulomb coupling. 

In the EPT, binary interactions occur via the potential of mean force
\begin{equation}
\label{eq:EPT}
\phi(r) = -k_{\B}T\ln[g(r)],
\end{equation}
which incorporates aspects of many-body correlations that become important at strong coupling~\cite{EPTRL,EPT}.
The potential of mean force is inserted in the momentum-transfer cross section by way of Eq.~(\ref{eq:scatangle}), and is modeled using the HNC theory of Eq.~(\ref{eq:OZ_HNC}) computed using \emph{Fozzie} \cite{Foz,Nathaniel_FozPap}.

EPT has also been extended to account for the effective exclusion volume about each particle due to Coulomb repulsion, which increases the collision frequency~\cite{ModEnskog}. 
To account for this effect, a  factor $\chi [ g(\eta/2)]$ is multiplied by the collision operator~\cite{ModEnskog}.
Here, $\eta$ defines the radius of the exclusion volume (in a hard sphere gas, this would be the diameter of the spheres). 
This was determined for the OCP by comparison with the virial expansion of the hard sphere equation of state predicted by Enskog's kinetic theory, as described in~\cite{ModEnskog}. 
The results for the coupling strengths of interest in this work ($\Gamma$ = 0.1, 1, 10, and 100) are $\chi$ = 1.02, 1.36, 1.45, and 1.65, respectively.

\section{Molecular Dynamics Simulations}
\label{sec:sims}

\subsection{Simulation details}

The energy evolution of charged projectiles (mass $m_\alpha$, charge $q$) evolving in the classical OCP is studied using MD simulations carried out using the \emph{CoulMD} code described in~\cite{Code}.
Simulations were initiated by first equilibrating an OCP background plasma to a fixed $\Gamma$ value using a velocity-scaling thermostat \cite{MDbook} in a periodic cubic domain.
In all cases, the thermostat was run for 500 $\omega_{p}^{-1}$.
A charged projectile was introduced with an initial velocity $\vc{V}_0$. 
The subsequent velocity evolution of the projectile $\vc{v}(t)$ was recorded. 
The length of the simulations varied depending on the plasma coupling strength and projectile mass. 
The charge of projectiles and background particles were all unity and of the same sign. 
Timesteps were chosen to be small enough to resolve high-speed collisions and the number of particles chosen so that the box size was much longer than the interaction length scale; specific values are provided in Table~\ref{tb:md}.
The wake of the projectile and the periodic boundary conditions of the simulation volume may potentially introduce unphysical artifacts \cite{Zwick_Review,MD_Grabowski}.
Convergence tests were carried out to confirm that the results did not depend on the box size, and spurious influences of the wake of the projectile with itself were negligible. 

A large statistical sample was obtained by repeating this procedure approximately 300 times at each set of conditions explored ($\vc{V}_0/v_T$, $\massrat=m_\alpha/m$, $\Gamma$).
However, the entire equilibration stage was not repeated each time. 
To obtain a statistically independent background plasma for each run at a fixed $\Gamma$, the state of the plasma at the beginning of the previous run was extended for 1 $\omega_p^{-1}$ with the thermostat on. 
A variety of coupling strengths ($\Gamma = 0.1, 1, 10, 100$), mass ratios $(\massrat = 1000, 10, 1)$ and initial velocities ($\vc{V}_0$) were explored.

\begin{table}
\label{tb:md}
\begin{ruledtabular}
\begin{tabular}{ccc}
$\Gamma$ & $N$ & $\Delta t$  $(\omega_{p}^{-1})$\\ [0.5ex] 
 \hline
 0.1 & $5.0 \times 10^{4}$ & $2.5 \times 10^{-3}$  \\ 
 1 & $1.0 \times 10^{4}$ & $1.0 \times 10^{-2}$  \\
 10 & $5.0 \times 10^{3}$ & $1.0 \times 10^{-2}$, $1.0 \times 10^{-3}$ \\
 100 & $5.0 \times 10^{3}$ & $1.0 \times 10^{-2}$, $1.0 \times 10^{-3}$ \\ [1ex] 
 \end{tabular}
\end{ruledtabular}
\caption{The number of background particles ($N$) and time step ($\Delta t$) used for each coupling strength ($\Gamma$) simulated.
Projectile speeds above 7 $v_{T}$ required a smaller time step, which is given as the second value in the 3$^{rd}$ column.}
\end{table}

\subsection{Projectile energy evolution}

\begin{figure*}
\centering
\includegraphics[width=15cm]{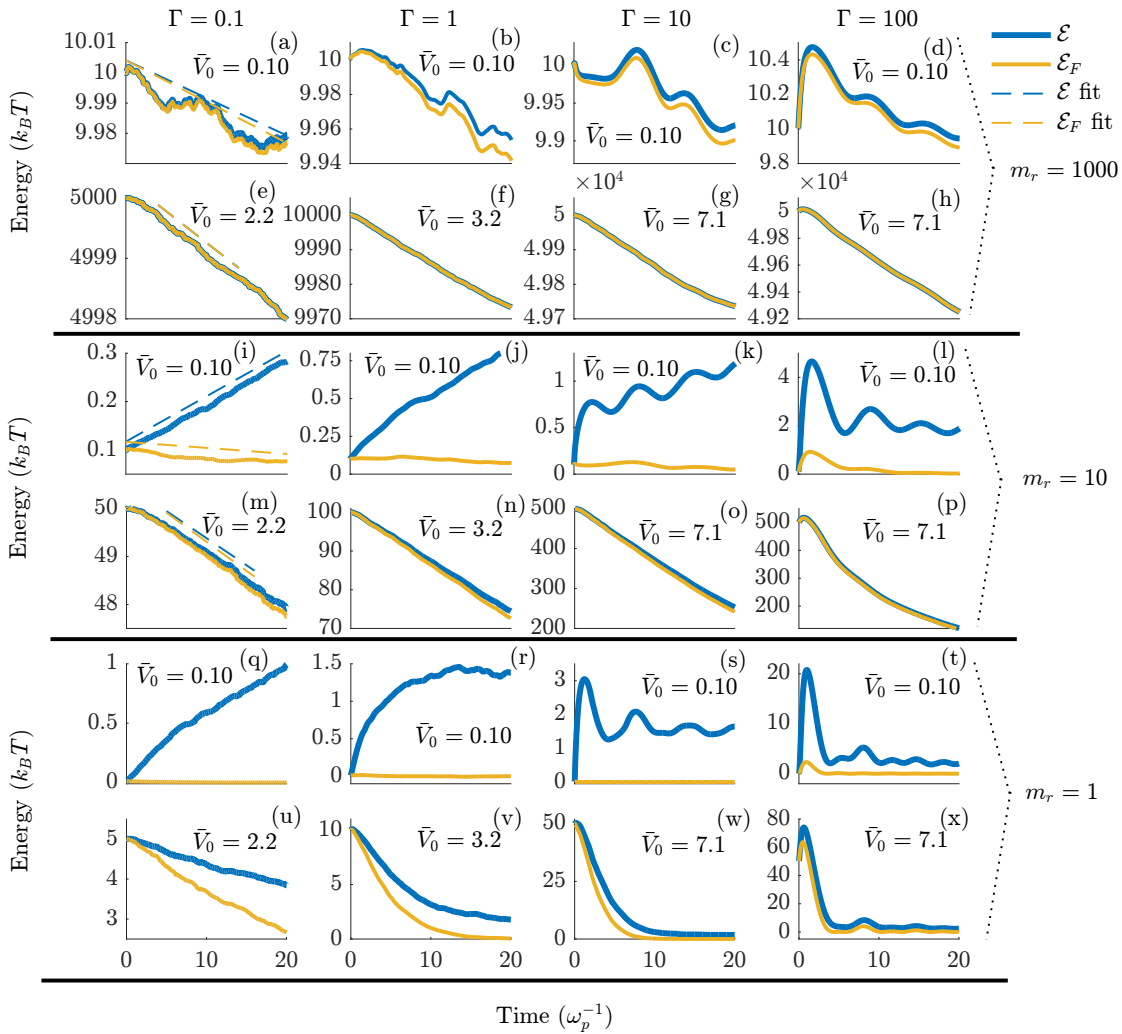}
\caption{Total projectile kinetic energy  $\mathcal{E}(t)$ (blue) and flow component $\mathcal{E}_\textrm{F}(t)$ (yellow) versus time for $\Gamma$ = 0.1, 1, 10, and 100 for example fast and slow projectile speeds. 
Panels (a)-(h), (i)-(p), and (q)-(x) show data for $\massrat$ = 1000, 10, and 1 respectively. 
The blue and and yellow dashed lines are examples of linear fits of $\mathcal{E}$ and $\mathcal{E}_F$, respectively.}
\label{fig:EvT}
\end{figure*}

Results for the time evolution of the average total projectile kinetic energy ($\mathcal{E}$) and the flow component of the kinetic energy ($\mathcal{E}_\textrm{F}$) are shown in Fig.~\ref{fig:EvT}. 
The component $\mathcal{E}_\textrm{T}$ of the kinetic energy can be inferred from $\mathcal{E}_\textrm{T} = \mathcal{E} - \mathcal{E}_\textrm{F}$. 
The conventional notion of stopping power described in Sec.~\ref{sec:conv} applies when the average kinetic energy is predominantly in the flow component ($\mathcal{E}_\textrm{F} \gg \mathcal{E}_\textrm{T}$) and the total average energy change is linear in time at early times. 
The more general description of stopping power as a two-component process, as described in Sec.~\ref{sec:tc}, requires only that the latter criterion be met. 
In addition, oscillations in each component of the average energy are observed at strong coupling, which are not part of the conventional stopping power models. 
Each of these features can be understood as follows. 
 
\subsubsection{Thermal contribution} 

If the initial kinetic energy of the projectile is smaller than the thermal energy of the background plasma, the projectile will gain more kinetic energy as it comes to thermal equilibrium with the background than it loses due to the slowing of its mean speed. 
Thus, the criterion for a small contribution from $\mathcal{E}_\textrm{T}$ to the total kinetic energy is expected to be $\frac{1}{2} m_\alpha V_0^2 \gg k_BT$. 
In the notation of Fig.~\ref{fig:EvT}, this criterion is $\bar{V}_0 \equiv  V_0/v_{\rm T} \gg 1/\sqrt{\massrat}$. 
If this criterion is met, it is expected that the conventional notion of 1D slowing, described in Sec.~\ref{sec:conv}, applies. 

This expectation is corroborated by the data shown in Fig.~\ref{fig:EvT}. 
For a large mass ratio (an example being fast fusion products such as an alpha particles slowing on plasma electrons), the total and flow components of the energy are nearly identical. 
In contrast, for $\massrat = 10$ this condition is met for fast projectiles ($\bar{V}_0 \gtrsim 1/\sqrt{\massrat} \simeq 0.3$), but not for slow projectiles. 
Panels (i)-(l) show that projectiles starting with an initial speed of $\bar{V}_0=0.1$ \emph{gain} kinetic energy in coming to equilibrium with the background. 
In these cases, the total stopping power is expected to be negative. 
These same trends are observed at unity mass ratio, where the thermal component of the projectile energy budget is significant for $\bar{V}_0 \lesssim 1$. 

\subsubsection{Linearity}

\begin{figure}
\includegraphics[width=7cm]{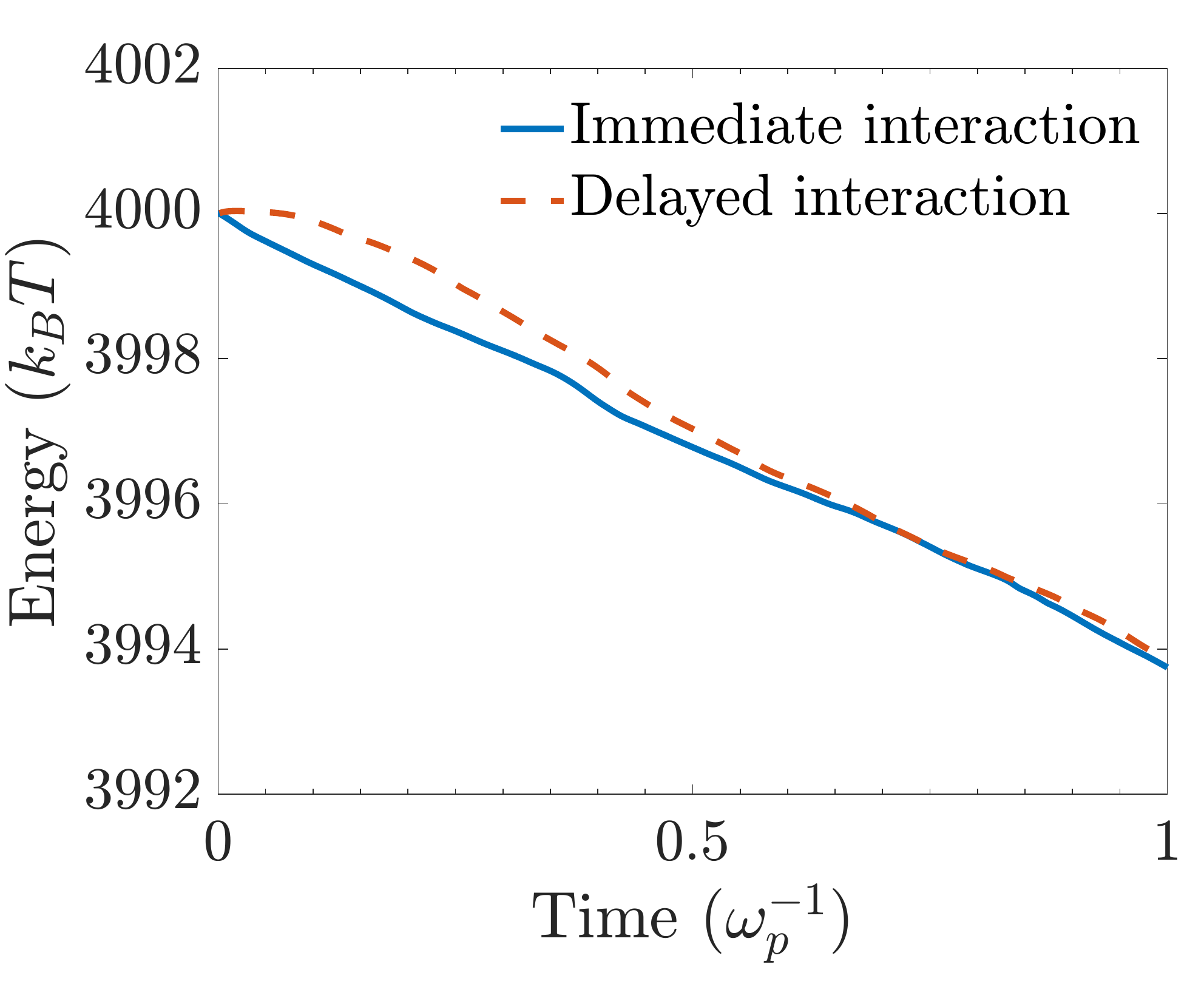}
\caption{Total kinetic energy evolution of a projectile with initial speed $\bar{V}_0 = 2$ in a plasma with $\Gamma = 1$. In one case the plasma interacted with the projectile immediately (dashed line), while in the other the projectile velocity was held fixed for a delay time of $20 \omega_p^{-1}$ before being influenced by the plasma (solid line). }
\label{fig:delay}
\end{figure}

The ordinary concept of stopping power to describe the energy evolution of a projectile applies when the projectile energy changes linearly in time, as indicated in Eq.~(\ref{eq:dEdx_dEdt}). 
In this case, stopping power is obtained from a linear fit to $\mathcal{E}(t)$ at an early time, divided by the initial speed. 
However, when a projectile is effectively placed at a random initial location (achieved by randomizing the background plasma in the simulations), its energy evolution is influenced by an initial transient occurring on the timescale of a few $\omega_p^{-1}$ (indicated by a flat region, or a kinetic energy increase, at early times). 
This arises due to placing the particle in a higher potential energy state than it would have evolved to if it was a part of the plasma at earlier times~\cite{Zwick_Transient}. 
This initial transient is not considered a part of the stopping power, and is not considered in any of the kinetic theories. 
One way to eliminate this initial transient effect in the simulations is to evolve the projectile at constant velocity for a fixed time after introducing it into the plasma, then subsequently turn off the constant velocity restriction and allow the projectile to interact with the plasma; as shown in Fig.~\ref{fig:delay}. 
Alternatively, since the initial transient is short, one can simply fit the region of the curve at slightly later times, after the transient effect is over; as shown in Fig.~\ref{fig:EvT}. 
Since the projectile kinetic energy change is very small over only a few $\omega_p^{-1}$ in cases where stopping power is meaningful, we find that either approach gives essentially indistinguishable results. 
We chose the latter option in obtaining stopping power from the energy evolution curves.

It was not always possible to identify a region of linear energy change at early times. 
In particular, if the projectile-plasma collision frequency was large enough that the energy changed significantly over the timescale of the initial transient, then a linear change was not observed. 
Similarly, at high Coulomb coupling, large-amplitude oscillations in the kinetic energy occur on a similar $\omega_p^{-1}$ timescale (a large rise and fall in kinetic energy, e.g. see Figs.~\ref{fig:EvT}(l) and (t)). 
Unless the Coulomb collision frequency is much smaller than $\omega_p$, there will be no clearly identifiable linear region of the energy evolution. 
In these cases, stopping power does not provide a meaningful representation of the energy evolution of the projectiles. 

\begin{figure}
\includegraphics[width=8cm]{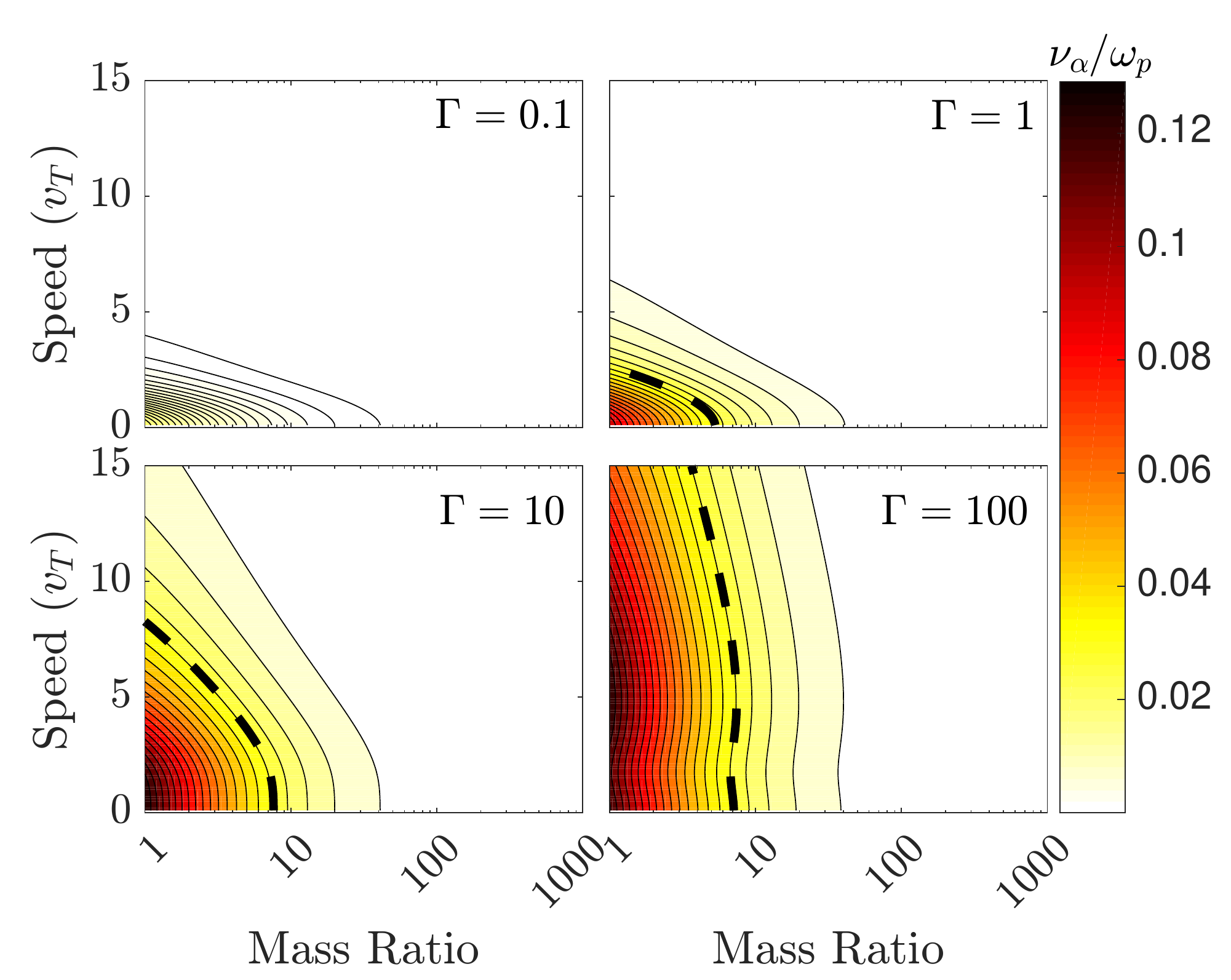}
\caption{Contours of constant Coulomb collision frequency ($\nu_{\alpha}/\omega_{p}$) as a function of projectile speed and mass ratio for $\Gamma$ values of 0.1, 1, 10 and 100. 
The dashed line marks $\nu_{\alpha}/\omega_{p} = 0.03$.}
\label{fig:Conts}
\end{figure}

Empirically, a linear energy evolution was observed when $\nu_{\alpha}/\omega_{p} \lesssim 0.03$. 
Figure~\ref{fig:Conts} shows contours of constant $\nu_{\alpha}/\omega_p$ computed from Eq.~(\ref{eq:collfreq}). 
This shows that at weak coupling ($\Gamma = 0.1$) a linear energy change (and hence stopping power) could be identified at all mass ratio and speeds explored. 
However, as the coupling strength increases, the conditions for a linear energy change are limited to sufficiently high mass ratios and sufficiently fast projectiles. 
Stopping power was computed only in cases where a linear energy change was observed ($\nu_{\alpha}/\omega_{p} \lesssim 0.03$). 

\subsubsection{Coulomb Coupling} 

Figure~\ref{fig:EvT} shows that qualitatively new features arise in the kinetic energy evolution at strong Coulomb coupling. 
Namely, oscillations in the kinetic energy (both total and flow components) are observed, particularly for low speed and low mass projectiles. 
By definition, the average potential energy of interacting particles exceeds their average kinetic energy at strong coupling. 
At these conditions, the continual exchange of potential and kinetic energy that arises due to correlations of the spatial positions of the background plasma are observable, and even dominant, features of the kinetic energy evolution. 
For example, at $m_\textrm{r} =1$ and $\bar{V}_0 = 0.1$ the initial rise and fall of the kinetic energy due to effectively placing the projectile in a random spatial location is the dominant feature of the kinetic energy evolution. 

One expects that these oscillations arise when the kinetic energy of the projectile is smaller than, or comparable to, the potential energy of its interaction with the background: $\frac{1}{2} m_\alpha V_0^2 \lesssim q^2/a$. 
In normalized units, this condition is $\bar{V}_0 \lesssim \sqrt{\Gamma/\massrat}$. 
Thus, even at strong coupling, the kinetic energy oscillations are small for sufficiently fast or massive projectiles. 
This expectation corresponds with the data shown in Fig.~\ref{fig:EvT}. 
Neither the notion of stopping power based on a linear energy change, nor the theories presented above, consider oscillations of the kinetic energy due to correlations.  
When computing stopping power based on a fit, these oscillations are averaged over, and the comparison to theory is based simply on the resulting linear slope. 

\section{Stopping Power}
\label{sec:Discussion}

\subsection{Large mass ratio} 

\begin{figure}
\includegraphics[width=8.5cm]{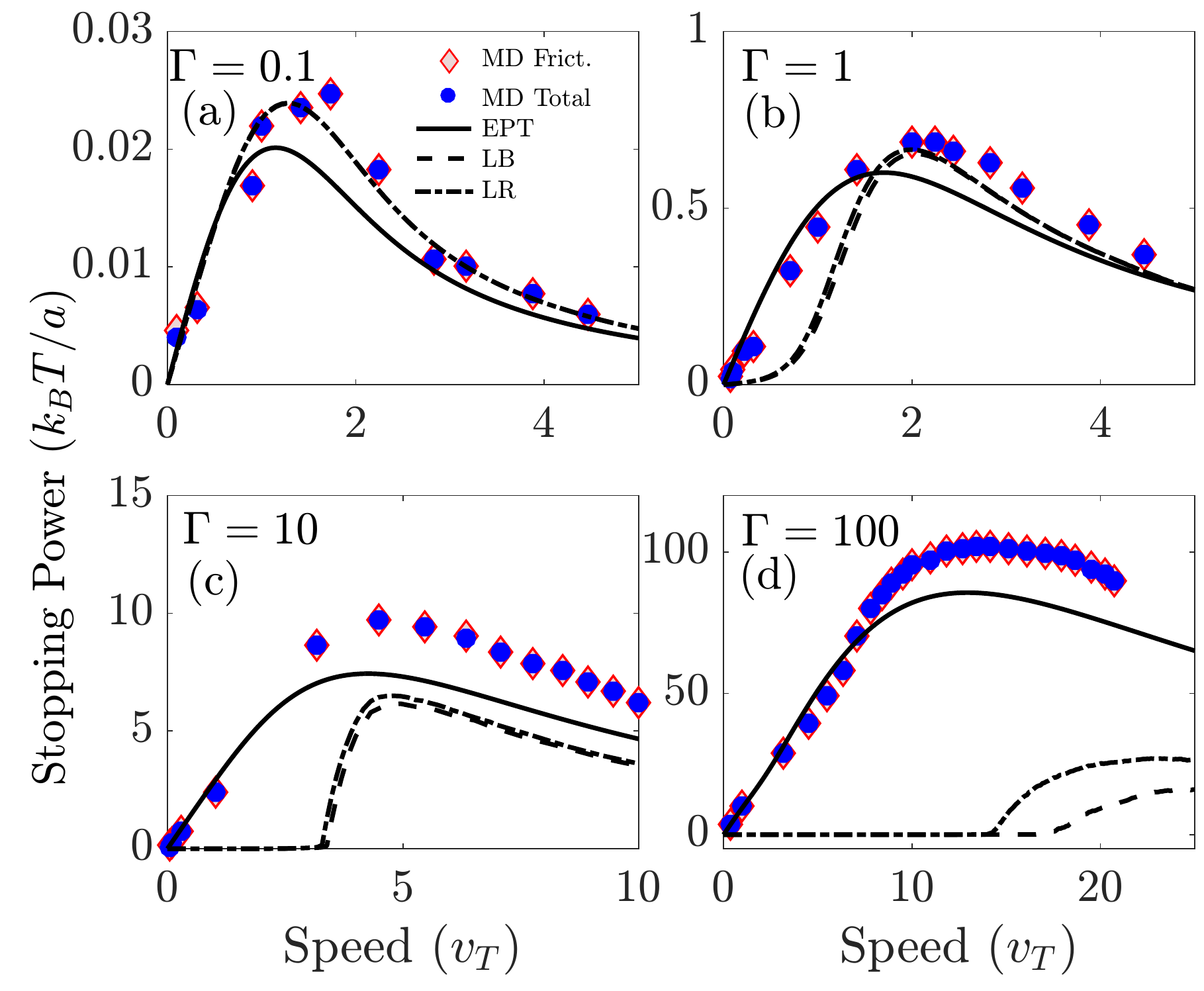}
\caption{Stopping power for $\massrat=1000$ at $\Gamma$ of 0.1, 1, 10, and 100: 
MD results for total (blue circles) and frictional (red diamonds) components, and theory curves from EPT [Eq.~(\ref{eq:Rept})] (solid lines), linear response with a local field correction (LR) [Eqs.~(\ref{eq:dEdx_dielectric}) and (\ref{eq:LFC_dielectric})] (dash-dotted line) and Lenard-Balescu collision operator (LB) [Eqs.~(\ref{eq:dEdx_dielectric}) and (\ref{eq:weakeps})] (dashed line). 
The dashed and dashed-dotted lines overlap in panel (a). 
}
\label{fig:dEdx_mr1000_all}
\end{figure}

The typical case of massive projectiles slowing on a light background is presented in 
Fig.~\ref{fig:dEdx_mr1000_all}, where $\massrat$ = 1000. 
Across coupling strengths, the MD simulation data show that there is little difference between the friction force component and the total stopping power, confirming that stopping power in this regime conforms to the conventional picture of one-dimensional slowing described in Sec.~\ref{sec:conv}.
As coupling increases, the speed at which the peak stopping power occurs (the Bragg peak) increases:  
The Bragg peak for $\Gamma =(0.1, 1, 10, 100)$ occurs at approximately $V_0/v_T = (1.7, 2.2, 4.5, 14)$.
The stopping power curve correspondingly broadens to span a larger range of speeds, and its magnitude in units of $k_\B T/a$ increases. 

A comparison with the theoretical models shows strengths and weaknesses of each. 
Since the stopping power is dominated by the friction component at large mass ratio, each of the theory curves is obtained from $d\mathcal{E}_\textrm{F}/dx$: Linear response (LR) from Eqs.~(\ref{eq:dEdx_dielectric}) and (\ref{eq:LFC_dielectric}), Lenard-Balescu (LB) from Eqs.~(\ref{eq:dEdx_dielectric}) and (\ref{eq:weakeps}), and EPT from Eq.~(\ref{eq:Rept}). 
At weak coupling ($\Gamma = 0.1$), all three models agree well at low speeds $V_0 \lesssim v_T$. 
Near the Bragg peak, the models based on dielectric response function (LR and LB) predict a slightly larger stopping power that is closer to the simulated value than is predicted by the collision model (EPT). 
This is a demonstration of the influence of ``dynamic screening,'' which is expected to contribute for fast projectiles and has been documented in previous studies~\cite{MD_Grabowski}. 

At stronger coupling, EPT continues to accurately predict the low-speed region of the stopping power curve; again extending nearly to the Bragg peak. 
Above the Bragg peak, it continues to under-predict the stopping power by a few tens of percent compared with MD. 
Again, this is likely associated with the neglect of dynamic screening in the collision model. 
In contrast, the models based on dielectric response begin to fail dramatically at low-speeds when the coupling is strong ($\Gamma \gtrsim 1$).  
If $\Gamma > 1$ and $V_0$ is sufficiently small, the dielectric models predict that $d\mathcal{E}_\textrm{F}/dx$ is inversely proportional to a positive power of $\Gamma$, resulting in near-zero stopping power values at low speeds (this is discussed in \cite{PeterMeyer,Arista_2018}). 
This prediction is inconsistent with the behavior observed in the simulations. 
Above the Bragg peak, these models predict a dramatic increase in the stopping power, which is in closer agreement with the simulations, but predicting a rate that is quantitatively much lower than the simulated value and even lower than the EPT prediction. 
Thus, we find that although the comparison of EPT and simulations suggests that dynamic screening should contribute above the Bragg peak, the existing linear response models are unable quantitatively predict the stopping power at strong coupling. 

\begin{figure}
\includegraphics[width=8cm]{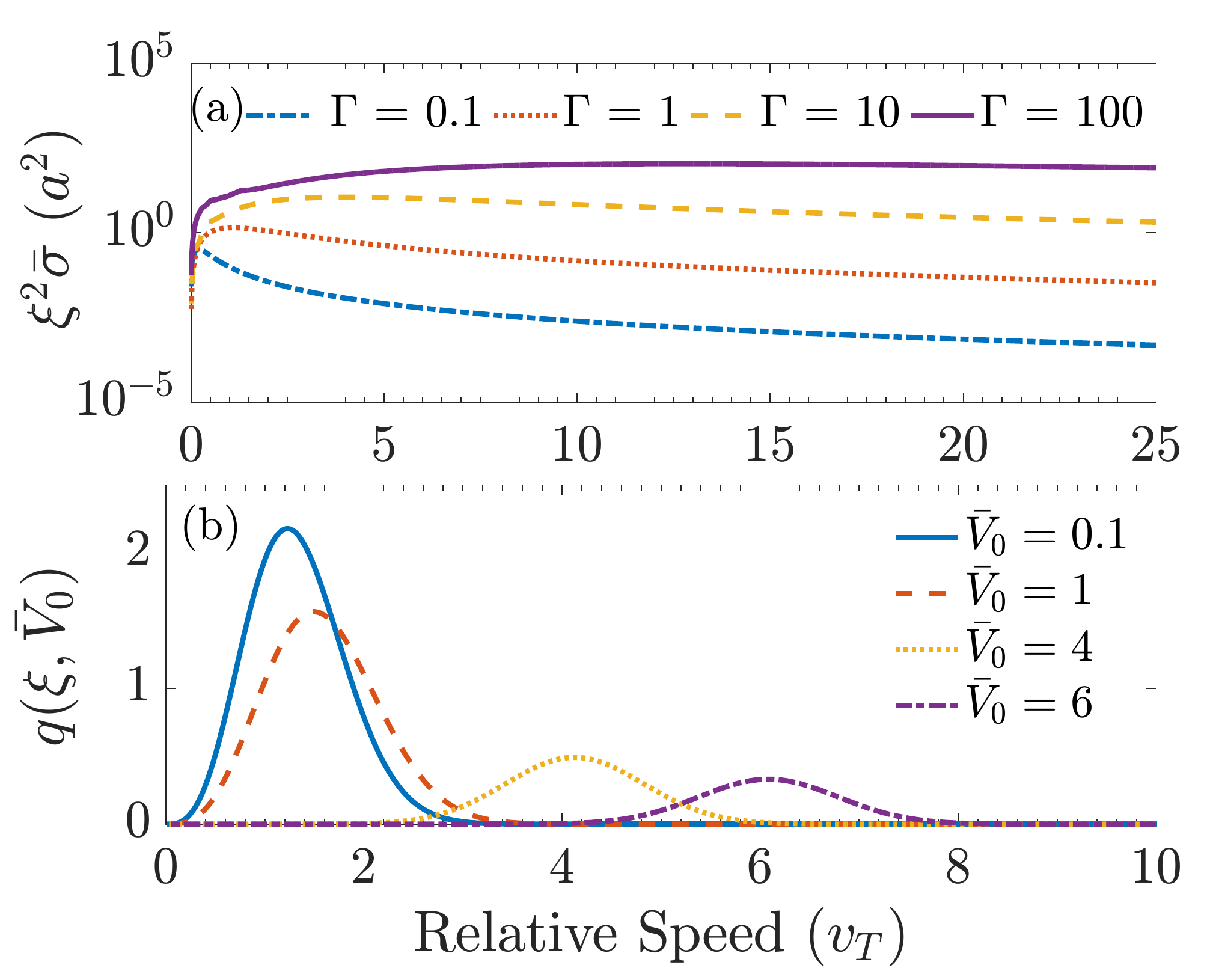}
\caption{Two terms in generalized Coulomb logarithm (Eq.~(\ref{eq:GenCoul})) integral computed for $m_\textrm{r} = 1000$. 
Panel (a) displays $\xi^{2} \bar{\sigma}^{(1)}(\xi)$ as a function of $\xi$ for $\Gamma$ = 0.1, 1, 10, and 100.
Panel (b) displays $q(\xi,\bar{V}_0)$ as a function of $\xi$ for $\bar{V}_0$ = 0.1, 1, 4, and 6.}
\label{fig:CrossSects_wx2}
\end{figure}

The shift of the Bragg peak, broadening of the curve, and increase in magnitude observed as coupling increases, can be qualitatively explained by dissecting the EPT model from Eqs.~(\ref{eq:Rept}) and (\ref{eq:GenCoul}).
The generalized Coulomb logarithm is composed of a term that describes the momentum transfer between the projectile and a background particle [$\xi^{2} \bar{\sigma}(\xi)$], and a term that describes which background particles are most likely to interact with the projectile at a given speed $[(2\xi \bar{V}_0+1) e^{-(\xi+\bar{V}_0)^2} + (2\xi \bar{V}_0 -1) e^{-(\xi-\bar{V}_0)^2}]/\bar{V}_0^{3}$, denoted here as $q(\xi,\bar{V}_0)$. 
Each of these terms are illustrated in Fig.~\ref{fig:CrossSects_wx2} for representative cases. 

\begin{figure*}
\includegraphics[width=15cm]{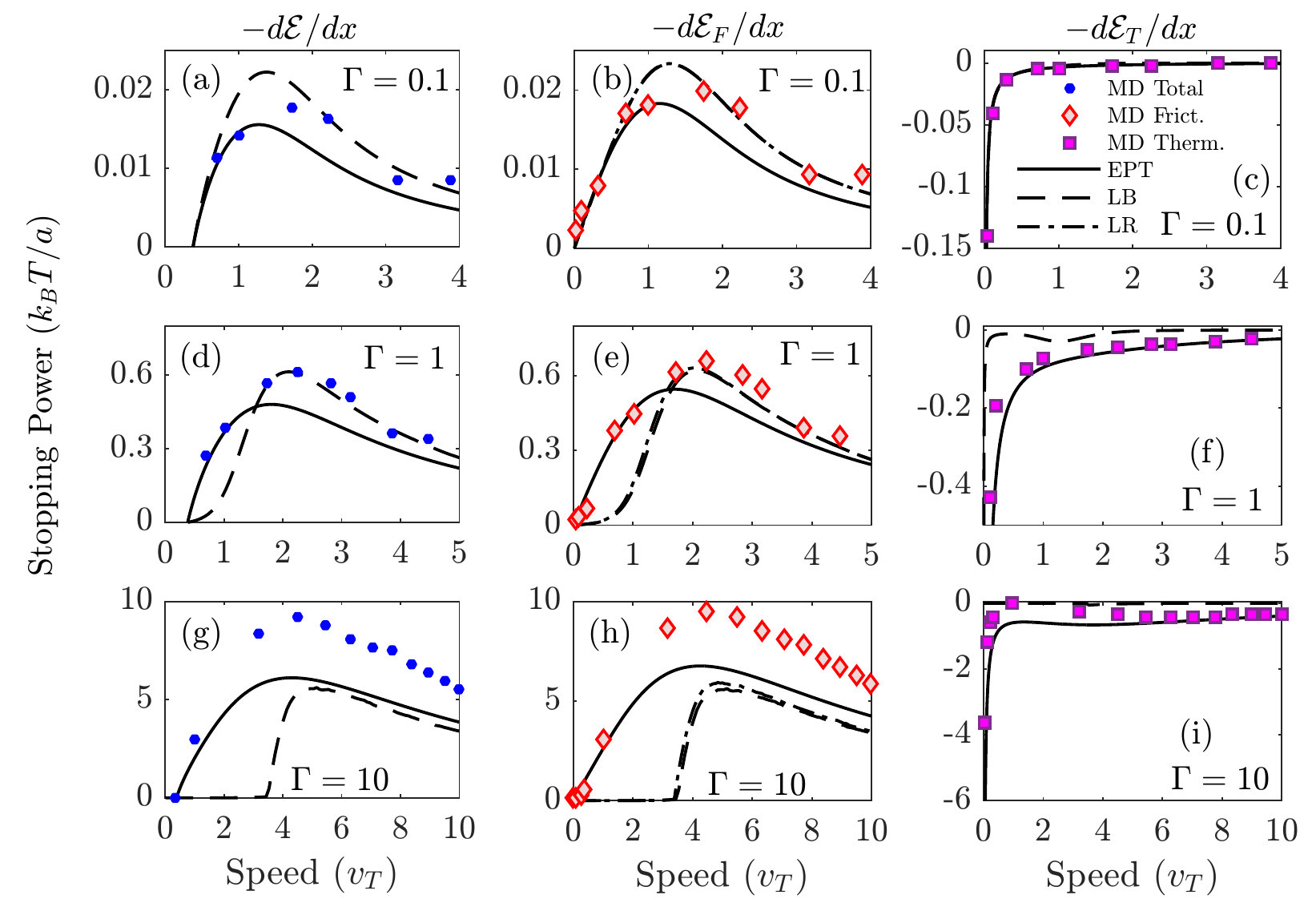}
\caption{Stopping power for $\massrat=10$ at $\Gamma$ of 0.1, 1, and 10. 
MD data is shown for each component of the stopping power: total $-d\mathcal{E}/dx$ (blue circles), frictional $-d\mathcal{E}_\textrm{F}/dx$ (red diamonds) and thermal $-d\mathcal{E}_\textrm{T}/dx$ (purple squares). 
Theoretical predictions are shown from each model for the frictional component: LR (dash-dotted lines), LB (dashed lines) and EPT (solid lines), and from the LB and EPT models for each of the other components. 
}
\label{fig:dEdx_mr10_all}
\end{figure*}

For low projectile speeds, the collision probability factor is independent of $\bar{V}_0$: $q(\xi, \bar{V}_0 \ll 1) \rightarrow \frac{16}{3} \xi^3 e^{-\xi^2}$. 
Since the cross section depends only on $\xi$, in this limit $\Xi^{[\textrm{F}]}_{\alpha}(V_0) \rightarrow \Xi$, where 
\begin{equation}
\label{eq:Vind_GenCoul}
\Xi = \frac{1}{2} \int_0^\infty d\xi \frac{\bar{\sigma}(\xi)}{\sigma_o} \xi^{5} e^{-\xi^2} 
\end{equation}
is independent of $\bar{V}_0$ and $d\mathcal{E}_\textrm{F}/dx \propto \bar{V}_0$ for $\bar{V}_0 \ll 1$; see Eq.~(\ref{eq:Rept}). 
For high projectile speeds, the most probable collisions are those with particles with speeds near the projectile speed $\bar{V}_0$ (the second exponential factor in $q(\xi, \bar{V}_0)$ dominates). 
Two factors contribute to a precipitous decline of $d\mathcal{E}_\textrm{F}/dx$ at large $\bar{V}_0$: There are fewer particles to collide with in the neighborhood of these higher energies and the cross section for such interactions also decreases at sufficiently high energy. 
For weak coupling, both the cross section and collision probability factor decrease when $\bar{V}_0\gtrsim 1$. 
For example, at weak coupling $\bar{\sigma}/\sigma_o \rightarrow 4\ln \Lambda/\xi^4$, and the stopping power decreases approximately as $\propto \bar{V}_0^{-2}$ for $\bar{V}_0 \gtrsim 1$~\cite{ConfProceedings}. 
The Bragg peak occurs at the intersection of the low and high speed limits. 

In contrast, at strong coupling the cross section factor $\xi^2 \bar{\sigma}$ is not sensitive to the relative speed until it is very high. 
This causes the Bragg peak to shift to higher projectile speed, and consequently broaden at strong coupling. 
At the same time, the stopping power increases in magnitude because the cross section is larger at strong coupling; see Fig.~\ref{fig:CrossSects_wx2}a. 
Microscopically, this is a consequence of collisions being predominately large angle. 
At strong coupling, screening limits the range of interactions to be amongst nearest neighbors, and those interactions have scattering angles of approximately $90^\circ$; see Fig.~5 of~\cite{EPT}. 
The relative speed between particles must be very fast in a strongly coupled plasma in order to have scattering angles much less than $90^\circ$. 
This is why the cross section is nearly constant until the relative speed of interacting particles becomes large in a strongly coupled plasma. 

\subsection{Intermediate mass ratio}

Figure~\ref{fig:dEdx_mr10_all} shows that for projectiles with a reduced mass ratio ($\massrat = 10$) stopping power consists of two contributions; one contribution is from one-dimensional slowing and the other is from thermalization with the background plasma, as described in Sec.~\ref{sec:tc}. 
At each coupling strength presented ($\Gamma = 0.1, 1$ and 10), the total stopping power is negative at sufficiently low projectile speed, indicating that the projectiles gains more kinetic energy in coming to equilibrium with the background plasma than they lose to friction. 
This situation corresponds to the linear kinetic energy gain shown in Figs.~\ref{fig:EvT}(i)-(k). 
Above the Bragg peak, the thermalization force ($d\mathcal{E}_\textrm{T}/dx$) only contributes slightly, making stopping power primarily a one-dimensional friction force ($d\mathcal{E}_\textrm{F}/dx$). 
Since the energy evolution was nonlinear for all projectile speeds at $\Gamma = 100$ at this mass ratio, stopping power was not well defined at this strong Coulomb coupling value; see Fig.~\ref{fig:Conts}. 

\begin{figure*}
\includegraphics[width=12cm]{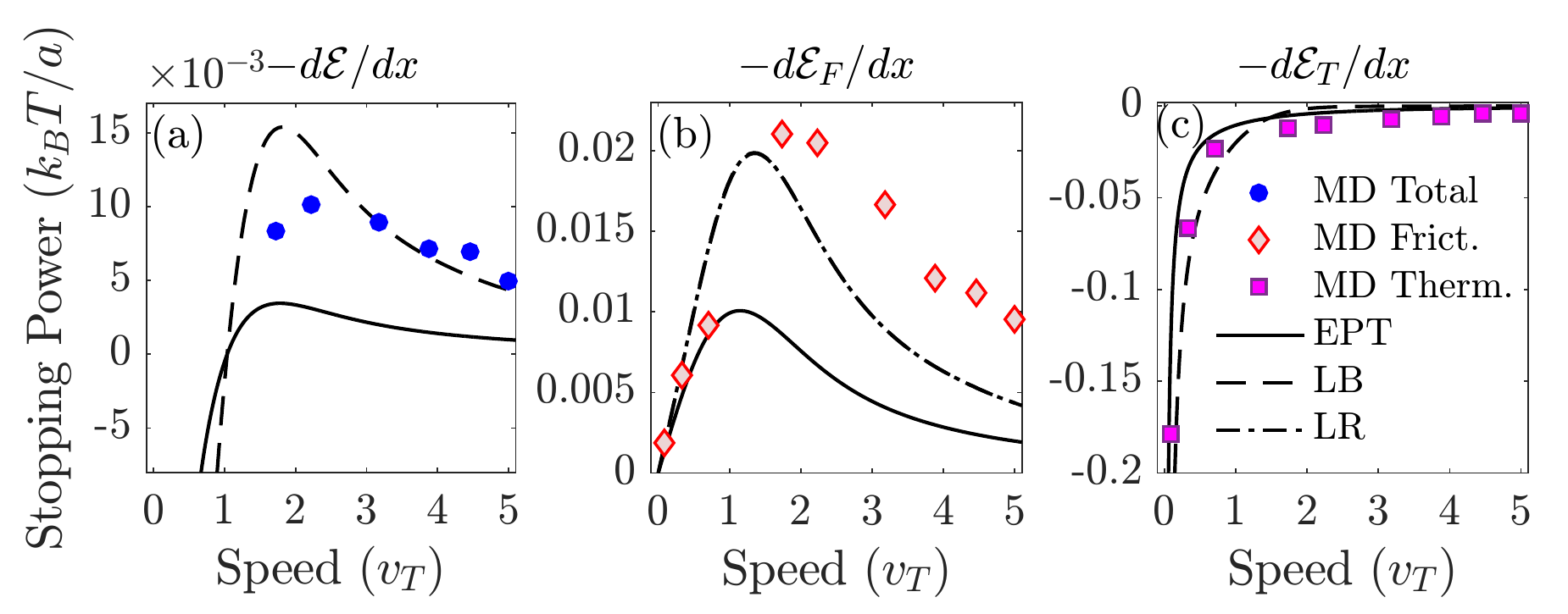}
\caption{Stopping power for $\massrat=1$ and $\Gamma = 0.1$.
MD data is shown for each component of the stopping power: total $-d\mathcal{E}/dx$ (blue circles), frictional $-d\mathcal{E}_\textrm{F}/dx$ (red diamonds) and thermal $-d\mathcal{E}_\textrm{T}/dx$ (purple squares). 
Theoretical predictions are shown from each model for the frictional component: LR (dash-dotted lines), LB (dashed lines) and EPT (solid lines), and from the LB and EPT models for each of the other components. 
}
\label{fig:dEdx_mr1_GP1}
\end{figure*}

Comparison with the theoretical models shows a similar trend as was observed at high mass ratio. 
The EPT predictions accurately model each component of the stopping power at low projectile speeds. 
At high projectile speeds (above the Bragg peak) it under-predicts the 1D frictional component by a few tens of percent. 
Again, this is presumably due to the lack of dynamic screening in this theory. 
The dielectric response models (both LR and LB) give nearly identical, and accurate, predictions for the frictional component at weak coupling ($\Gamma = 0.1$), but fail for low speeds when $\Gamma > 1$ in a similar way as at large mass ratio. 
Additionally, the LB model from Eq.~(\ref{eq:dEdx_dieltherm}) provides a prediction for the thermalization term, and hence the total stopping power. 
Again, this prediction is found to be accurate at weak coupling ($\Gamma = 0.1$), but to quickly break down for $\Gamma \gtrsim 1$ at low speeds, where the thermalization term contributes. 

Figure~\ref{fig:dEdx_mr1_GP1} shows that similar trends continue at unity mass ratio $\massrat =1$. 
The thermal contribution to the stopping power is very large at low speeds in this case, driving the total stopping power to a large negative value for $\bar{V}_0\lesssim 1$. 
The contribution of the thermal term continues to be significant in comparison to the friction component even well above the Bragg peak. 
Again, the EPT model is accurate at low projectile speeds, and the LB model is more accurate at high projectile speeds. 
However, both models seem to significantly under-predict the 1D friction component at high projectile speeds at this mass ratio. 
Only the weakly coupled case is shown here because the energy evolution was observed to be nonlinear at all other $\Gamma$ values tested; see Figs.~\ref{fig:EvT} and \ref{fig:Conts}. 
These reduced mass ratio results are particularly relevant to very slow ions in a plasma that primarily stop due to collisions with other ions~\cite{PeterMeyer}, or to muons stopping on electrons, which have been proposed as a method to catalyze fusion reactions~\cite{muon}.

\section{Relating Stopping Power to Macroscopic Transport Properties} 

\subsection{Diffusion}

A useful connection can be made between the frictional component of the stopping power of a massive projectile and the near-equilibrium property of the self-diffusion coefficient of the background plasma~\cite{Dufty1,Dufty2}. 
This quantity is also related to the mutual diffusion coefficient for a massive impurity species and a background plasma. 
This connection may provide an avenue to effectively measure a macroscopic transport property based on a microphysical process, which is especially beneficial in dense plasma experiments where macroscopic transport properties are generally difficult to measure, but where accurate methods for measuring stopping power have already been demonstrated (e.g. ~\cite{Experiment_WDM,Experiment_BraggPeak,Slow_Exp,SlowParts_Exp}). 

\begin{figure*}
\includegraphics[width=15cm]{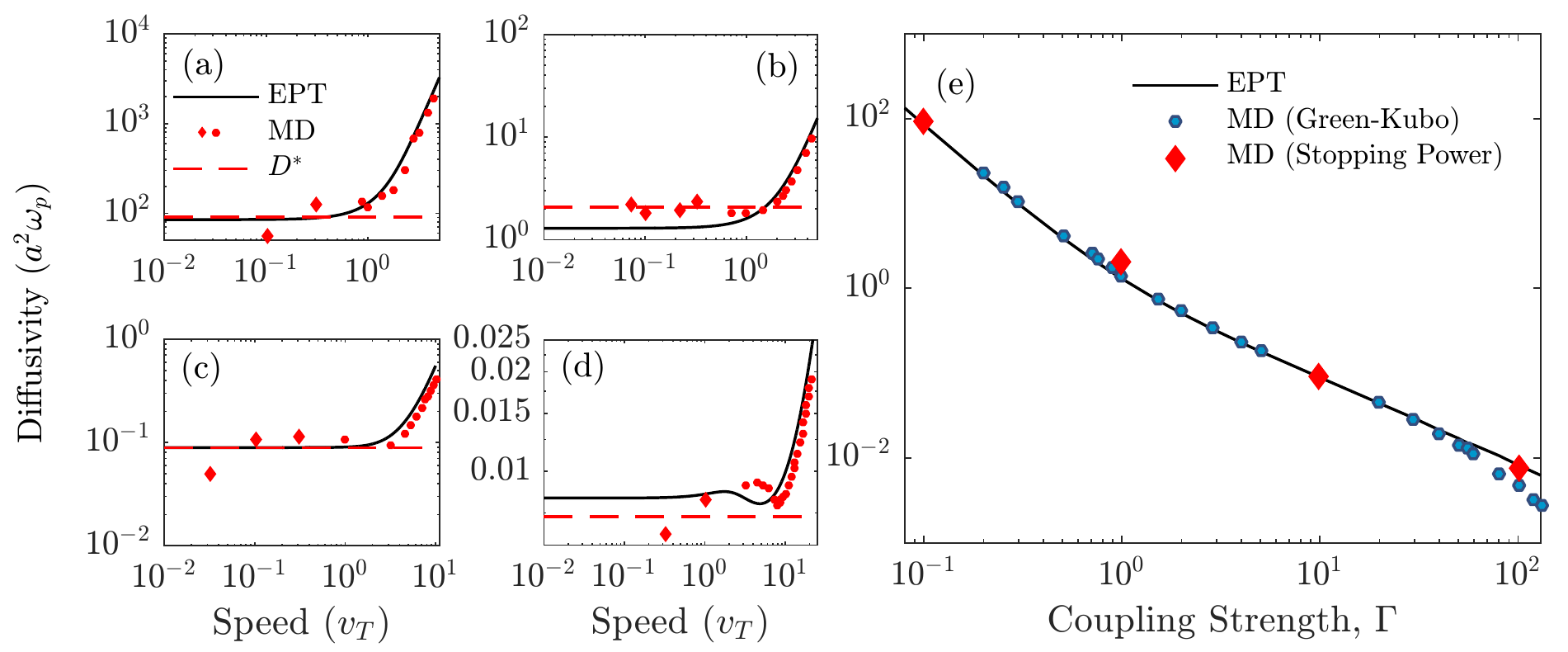}
\caption{Comparison between MD results for velocity dependent Eq.~(\ref{eq:Diffusion}) and EPT for: (a) $\Gamma = 0.1$, (b) $\Gamma=1$, (c) $\Gamma = 10$, and (d) $\Gamma = 100$. Diamonds are data points averaged over to obtain diffusion coefficients, and the red dashed line marks the obtained value. (e) Diffusion coefficients extracted from the MD simulations (red diamonds), MD simulations based on the Green-Kubo method from \cite{VAF} (circles), and predictions of EPT. \label{fig:Diff_vs_V} }
\end{figure*}

Dufty has proposed a general connection between these quantities based on the velocity autocorrelation function~\cite{Dufty1,Dufty2}. 
Here, we test a similar connection from the EPT model. 
The mutual diffusion coefficient obtained from applying the first-order Chapman-Enskog solution to the EPT kinetic equation is~\cite{shaf:17} 
\begin{equation}
\frac{[D_{\alpha}]_1}{a^2 \langle \omega_p \rangle} = \sqrt{\frac{\langle m \rangle}{2\mu}} \frac{\sqrt{\pi/3}}{\Gamma^{5/2}} \frac{1}{\Xi_{\alpha}}
\end{equation} 
where $\langle m \rangle = (n_\alpha m_\alpha + n m)/(n_\alpha+n)$, $n_\alpha$ is the density of species $\alpha$, $\langle \omega_p \rangle = \sqrt{4\pi ne^2/\langle m \rangle}$, and $\Xi_{\alpha}$ is the multispecies generalized Coulomb logarithm in the low speed limit~\cite{EPT}. 
In the limit of a massive impurity ($n_\alpha/n \rightarrow 0$ and $m_\alpha \gg m$), the mass dependence in this expression is determined solely by the background plasma $\mu = m$ and $\langle m \rangle = m$, so $\langle \omega_p \rangle \rightarrow \omega_p$ and $\Xi_{\alpha} \rightarrow \Xi$ [from Eq.~(\ref{eq:Vind_GenCoul})].  
This can then be connected to the EPT prediction for stopping power from Eq.~(\ref{eq:Rept}) in the limit $\bar{V}_0 \rightarrow 0$ 
\begin{equation}
\label{eq:Diffusion}
\frac{[D_{\alpha}]_1}{a^2 \langle \omega_p \rangle} \rightarrow \frac{1}{\sqrt{2}} \frac{[D]_1}{a^2\omega_{p}} = \lim_{V_0\rightarrow 0} \frac{k_{B}T}{a^2\omega_{p}}\frac{V_0}{d\mathcal{E}_\textrm{F}/dx},
\end{equation}
where $[D]_1/(a^2\omega_p) = \sqrt{\pi/3}/(\Gamma^{5/2} \Xi)$ is the first-order Chapman-Enskog solution for the self-diffusion coefficient of a one-component plasma~\cite{EPT}. 

We note that the complete diffusion coefficient, described for example by the Green-Kubo relation, includes the influence of the distortion of the distribution functions from equilibrium due to the force (concentration gradient) causing the diffusive flux (mixing). 
This can be accounted for in the higher order terms of the Chapman-Enskog solution. 
However, this physics is not a part of the connection with stopping power because the background plasma in the case of stopping power is at equilibrium, and the ``distribution'' of the other species is a single projectile. 
Thus, there is a slight disconnect between Eq.~(\ref{eq:Diffusion}) and a full macroscopic diffusion coefficient. 
Nevertheless, it is known that this higher-order effect is small in weakly coupled plasmas (of the order of $20\%$) and is even smaller at strong coupling (essentially negligible); see for example Fig.~9 of \cite{EPT}. 

Figure~\ref{fig:Diff_vs_V} shows a comparison of Eq.~(\ref{eq:Diffusion}) with MD simulations for $\massrat =1000$ and four coupling strengths. 
This generally shows good agreement as the projectile speed asymptotes toward zero. 
The scatter in the MD data increases at low speeds, which hinders the comparison somewhat, especially at strong coupling. 
Panel (e) shows a comparison of the resulting diffusion coefficients with theory predictions across a broad range of coupling strength. 
Here, the MD data points were taken from fits to the low speed region of the curves shown in panels (a)-(d). 
This panel also includes MD data for the self-diffusion coefficient of an OCP from \cite{VAF}, which was computed using the Green-Kubo relation (i.e., by integrating the velocity correlation function). 

The good agreement demonstrates that the connection with stopping power provides an alternative way to measure a macroscopic transport property via the microphysical process of the slowing of individual projectiles. 
This suggests that measurements of the low-speed region of a stopping power curve (below the Bragg peak)~\cite{SlowParts_Exp} may also be used to infer the self-diffusion coefficient or the mutual diffusion coefficient of an impurity species. 

\subsection{Temperature relaxation}

\begin{figure*}
\includegraphics[width=15cm]{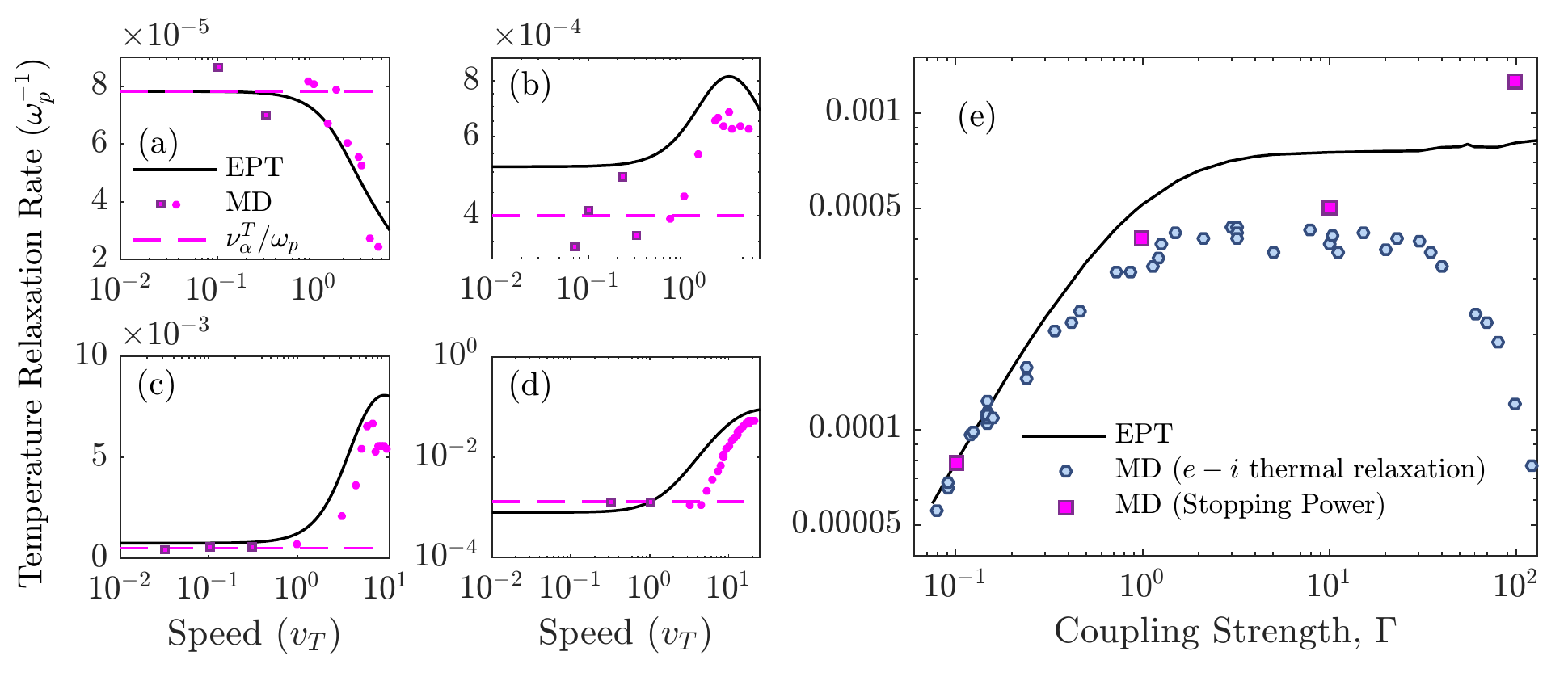}
\caption{Comparison between MD results for velocity dependent Eq.~(\ref{eq:nu_dEdx}) and predictions from EPT for: (a) $\Gamma = 0.1$, (b) $\Gamma=1$, (c) $\Gamma = 10$, and (d) $\Gamma = 100$.
Squares are data points averaged over to obtain temperature relaxation rates, and the purple dashed line marks the value of the average.
(e) Temperature relaxation rate extracted from these MD simulations (purple squares), previous MD simulations of macroscopic temperature relaxation~\cite{Code} (circles) and predictions from EPT.} 
\label{fig:Qs_vs_V}
\end{figure*}

Analogous to the connection between diffusion coefficients and the low-speed limit of $d\mathcal{E}_\textrm{F}/dx$, temperature relaxation rates can be related to the low-speed limit of $d\mathcal{E}_\textrm{T}/dx$ for a massive projectile. 
This connection is particularly relevant to the electron-ion temperature relaxation rate~\cite{Code}. 
Within the EPT model, the temperature relaxation rate between two Maxwellian distribution functions of slightly differing temperatures is~\cite{EPT}
\begin{equation}
\label{eq:dTdt}
\frac{dT_\alpha}{dt} =- \nu^T_{\alpha} (T_\alpha - T)
\end{equation}
where 
\begin{equation}
\frac{\nu_{\alpha}^T}{\langle \omega_p \rangle} = \frac{4}{\sqrt{6\pi}} \frac{m_\alpha}{m} \biggl( \frac{\mu}{m_\alpha} \biggr)^{3/2} \biggl( \frac{\langle m \rangle}{m_\alpha} \biggr)^{1/2} \Gamma^{3/2} \Xi_{\alpha}
\end{equation} 
and it is assumed that $T_\alpha \approx T$ so $\Gamma$ is unambiguously defined. 
In the limit that one of these species is a massive impurity, this temperature relaxation rate can be related to the generalized Coulomb logarithm from Eq.~(\ref{eq:Vind_GenCoul}), and in turn to $d\mathcal{E}_\textrm{T}/dx$ via Eq.~(\ref{eq:Qept}) 
\begin{subequations}
\begin{align}
\label{eq:nu_t_ept}
\frac{\nu_{\alpha}^T}{\langle \omega_p \rangle} \rightarrow \frac{\nu_{\alpha}^T}{\omega_p} &= \frac{4}{\sqrt{6\pi}} \frac{m}{m_\alpha} \Gamma^{3/2} \Xi \\ 
\label{eq:nu_dEdx}
&= \lim_{V_0\rightarrow 0} \frac{2}{3} \frac{V_0 d\mathcal{E}_\textrm{T}/dx}{\omega_p k_BT} .
\end{align}
\end{subequations} 
Although Eq.~(\ref{eq:dTdt}) is the temperature relaxation rate between two Maxwellian distribution functions of slightly different temperature, the temperature of the massive species does not contribute to the expression for the rate $\nu_{\alpha}^T$ as long as $m_\alpha \gg m$. 
Thus, the macroscopic temperature relaxation rate can be connected to the thermalization rate of single particle in the massive impurity limit. 

Figure~\ref{fig:Qs_vs_V} shows a comparison between MD simulations and EPT solutions for Eq.~(\ref{eq:nu_dEdx}) at $\massrat = 1000$. 
The agreement is similar to what was observed for the diffusion coefficient. 
Even at a large mass ratio, the asymptotic value is reached when the projectile speed is just below the Bragg peak. 
Since the MD data for this quantity is obtained via the subtraction $d\mathcal{E}_\textrm{T}/dx = d \mathcal{E}/dx - d\mathcal{E}_\textrm{F}/dx$, the statistical variation is somewhat larger than for $d\mathcal{E}_\textrm{F}/dx$ directly. 
Panel (e) shows a comparison between the temperature relaxation values obtained from the MD data in the limit $\bar{V}_0 \rightarrow 0$ and EPT solutions of Eq.~(\ref{eq:nu_t_ept}). 
This figure also shows the electron-ion temperature relaxation rate from~\cite{Code}, which is a direct MD simulation of the relaxation rate of near-thermal electron and ion distributions. 

The good agreement demonstrates that the connection with stopping power also provides an alternative way to measure temperature relaxation rates via a microphysical process. 
This suggests that measurements of low-speed stopping~\cite{SlowParts_Exp} may also be used to infer temperature relaxation rates. 
However, a potential challenge in this regard is that the interpretation of stopping power measurements typically assumes that the conventional one-dimensional slowing described in Sec.~\ref{sec:conv} is obeyed (i.e., that all projectiles slow in one-dimension along their initial velocity vector). 
In contrast, the thermalization term has to do with the statistical spread of velocities about this mean. 
It may require new diagnostic techniques, or new methods for interpreting the data using current techniques, in order to access $d \mathcal{E}_\textrm{T}/dx$ experimentally.

\subsection{Inferences from EPT} 

Section~\ref{sec:Discussion} established that the EPT model accurately predicts stopping power over a broad range of coupling strengths, particularly for projectile speeds below the Bragg peak. 
Since each component of the stopping power in this model is proportional to a Coulomb collision frequency, the relative importance of each contribution can be described in terms of the ratio of the associated generalized Coulomb logarithm terms.
Using Eqs.~(\ref{eq:Rept}) and (\ref{eq:Qept}), we note that the total stopping power can be expressed as
\begin{equation}
\frac{d\mathcal{E}}{dx} = \frac{d \mathcal{E}_\textrm{F}}{dx} \biggl( 1 - \frac{3}{2} \frac{\mu}{m_\alpha} \frac{1}{\bar{V}_0^2} \frac{\Xi^{[\textrm{T}]}_{\alpha}}{\Xi^{[\textrm{F}]}_{\alpha}} \biggr)
\end{equation}
where the first term in parenthesis is due to the 1D slowing, and the second is due to thermalization. 
We note that in the slow projectile limit $\lim_{\bar{V}_0 \rightarrow 0} \Xi^{[\textrm{T}]}_{\alpha}/\Xi^{[\textrm{F}]}_{\alpha} \rightarrow 1$. 
In the massive impurity limit, $\mu/m_\alpha \approx m/m_\alpha \ll 1$, and the second term is generally small unless the projectile velocity is very small ($\bar{V}_0\lesssim \sqrt{m/m_\alpha}$). 
This is simply an expression of the fact that energy exchange occurs on a slower timescale than momentum exchange, by a factor of approximately the mass ratio. 
Thus, the speed below which the thermalization contribution becomes significant is determined by the mass ratio. 

\begin{figure}
\includegraphics[width=8.0cm]{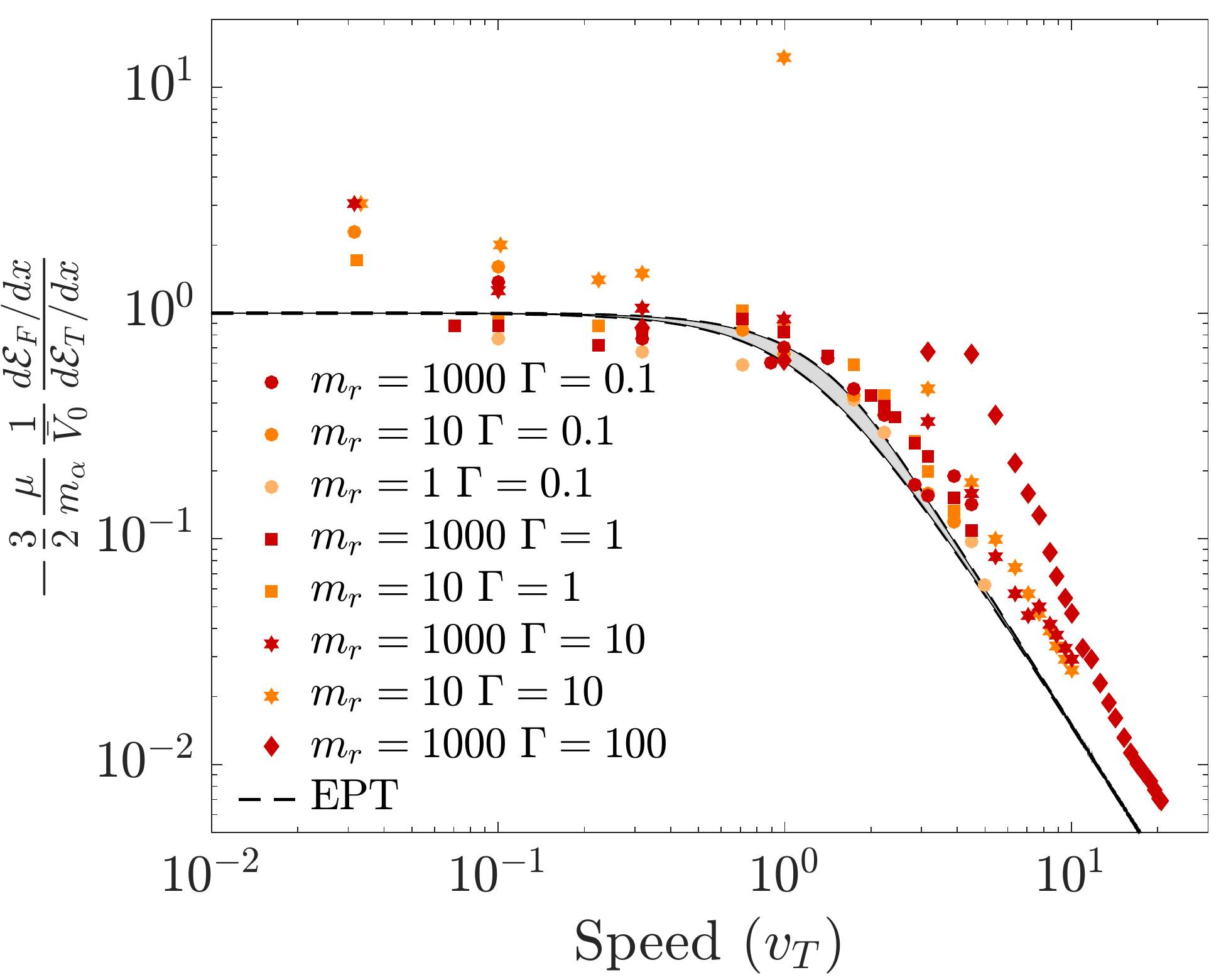}
\caption{Comparison of the prediction of Eq.~(\ref{eq:ratio}) computed from EPT and an aggregate of all MD simulation data. The shaded region indicates the range of EPT predictions over the range of $\Gamma$ values. }
\label{fig:ratio}
\end{figure}

EPT also predicts that the ratio $\Xi^{[\textrm{T}]}_{\alpha}/\Xi^{[\textrm{F}]}_{\alpha}$ does not depend on the mass ratio. 
This prediction can be tested from an aggregate of the MD data. 
In particular, Fig.~\ref{fig:ratio} shows a plot of the quantity
\begin{equation}
\label{eq:ratio}
- \frac{3}{2} \frac{\mu}{m_\alpha} \frac{1}{\bar{V}_0^2} \frac{d\mathcal{E}_\textrm{F}/dx}{d\mathcal{E}_{\textrm{T}}/dx} = \frac{\Xi^{[\textrm{F}]}_{\alpha}}{\Xi^{[\textrm{T}]}_{\alpha}}
\end{equation}
in which the left side is computed from MD and the right side from EPT. 
This demonstrates that the data at each mass ratio and coupling strength asymptotes to a universal value in the small speed limit, in accordance with the prediction. 
There is a slight $\Gamma$ dependence at large speeds that is predicted. 
It is also observed that the $\Gamma = 100$ data is furthest from the EPT prediction, but this is a coupling strength regime beyond which the EPT model is known to be accurate~\cite{EPT}. 

\section{Conclusions}
\label{sec:conc}

MD simulations were used to show the limitations, as well as extensions, of the conventional notion of stopping power. 
Stopping power accurately describes the energy evolution of a charged projectile when the energy of that projectile changes approximately linearly in time. 
Two effects were found to be required to satisfy this condition.
One was that the Coulomb collision frequency must be sufficiently small compared to the plasma frequency $\nu_{\alpha } \lesssim 10^{-2} \omega_p$. 
When this condition was violated, the projectile kinetic energy decay was inseparable from initial transients so an initial decay rate could not be identified. 
The second condition arose at strong coupling. 
When the initial kinetic energy of the projectile was smaller than the average potential energy of its interaction with the background plasma ($\frac{1}{2}m_\alpha V_0^2 \lesssim q^2/a$) large oscillations in the projectile kinetic energy were observed as it traversed the potential energy landscape associated with the background plasma. 

In the regimes where stopping power could be identified from a linear energy decay rate, it was found that Coulomb coupling tends to increase the magnitude of the stopping power, cause the Bragg peak to shift to higher speed (in terms of the thermal speed of the background plasma), and to cause the stopping power curve to broaden as a result. 
It was also shown that the total energy decay rate can be split into two components: one associated with the 1D slowing of the projectile along its initial velocity vector, and another associated with the thermalization of the projectile with the background plasma. 
This association was shown to resolve a discrepancy between a collisional stopping power model derived in \cite{Binary_FA} and the traditional models derived from linear-response theory.
The contribution from thermalization is not typically considered in stopping power models, but was shown to be significant for sufficiently slow projectiles ($\bar{V}_0 \lesssim \sqrt{m/m_\alpha}$). 

Stopping power curves were compared with theoretical predictions from the leading models. 
It was found that those based on dielectric response theory were typically accurate at weak coupling, and for fast projectile speeds (above the Bragg peak).  
However, these models break down at strong coupling, predicting a much lower stopping power for speeds below the Bragg peak, and generally under-predicting the stopping power at all speeds when the coupling strength was large. 
Models based on Coulomb collisions were found to be similarly accurate for low speed projectiles (below the Bragg peak) at weak coupling, but to slightly under-predict the stopping power for speeds above the Bragg peak. 
The EPT model was found to provide an accurate extension to strong coupling, with the trend that the model was particularly accurate for low speeds, and slightly under-predicted (by approximately 20\%) the stopping power of fast projectiles. 

The low-speed region of the stopping power curve was related to macroscopic transport rates. 
In particular, it was shown that the frictional component of the stopping power of a massive projectile ($d\mathcal{E}_\textrm{F}/dx$) is related to the self-diffusion coefficient of the background plasma. 
Similarly, the thermalization component of the stopping power of a massive projectile ($d\mathcal{E}_\textrm{T}/dx$) is related to the temperature relaxation rate. 
This connection was validated by comparing the rates computed from the stopping power curves with independent near-equilibrium MD methods from previous literature~\cite{Code,VAF}. 
The good agreement suggests that experimental measurements of low-speed stopping power may be used to also infer macroscopic transport rates of the plasma~\cite{Slow_Exp,SlowParts_Exp}.  

\appendix
\section{Models for the Dynamic LFC\label{sec:DLFC}}

Two models for the dynamic LFC $G(k,\omega)$ are considered here.
The first is that proposed by Ichimaru \emph{et al}~\cite{Ich_DLFC} 
\begin{equation}
\label{eq:Ich_DLFC}
G_\textrm{I}(k,\omega) = \frac{\omega G(k,\omega \rightarrow \infty) +i \omega G_\textrm{S}(k)}{\omega + i \omega_p},
\end{equation}
where $G(k,\omega \rightarrow \infty) = 2I(k)$ and
\begin{equation}
\label{eq:Ik}
I(k) = \int_0^\infty \frac{[ g(r) -1]}{r}\Bigg[\frac{\sin(kr)}{kr}+\frac{3\cos(kr)}{(kr)^2}-\frac{3\sin(kr)}{(kr)^3}\Bigg] dr;
\end{equation}
see Ref.~\cite{HMP_Ik}.
We also note that a similar result to Eq.~(\ref{eq:Ich_DLFC}) was derived from a generalized viscoelastic formalism in \cite{Ich_visco_DLFC}. 
The second model is that proposed by Hong and Kim~\cite{HK_DLFC} 
\begin{multline}
\label{eq:HK_DLFC}
G_\textrm{HK}(k,\omega) = G_\textrm{S}(k) - \frac{1}{2}[G_\textrm{S}(k) - G(k,\omega \rightarrow \infty)]  \\ \times \Big[\frac{-2}{Z'(\omega/v_Tk)}+2\Big(\frac{\omega}{kv_T}\Big)^2-1 \Big].
\end{multline}

\begin{figure}
\includegraphics[width=6cm]{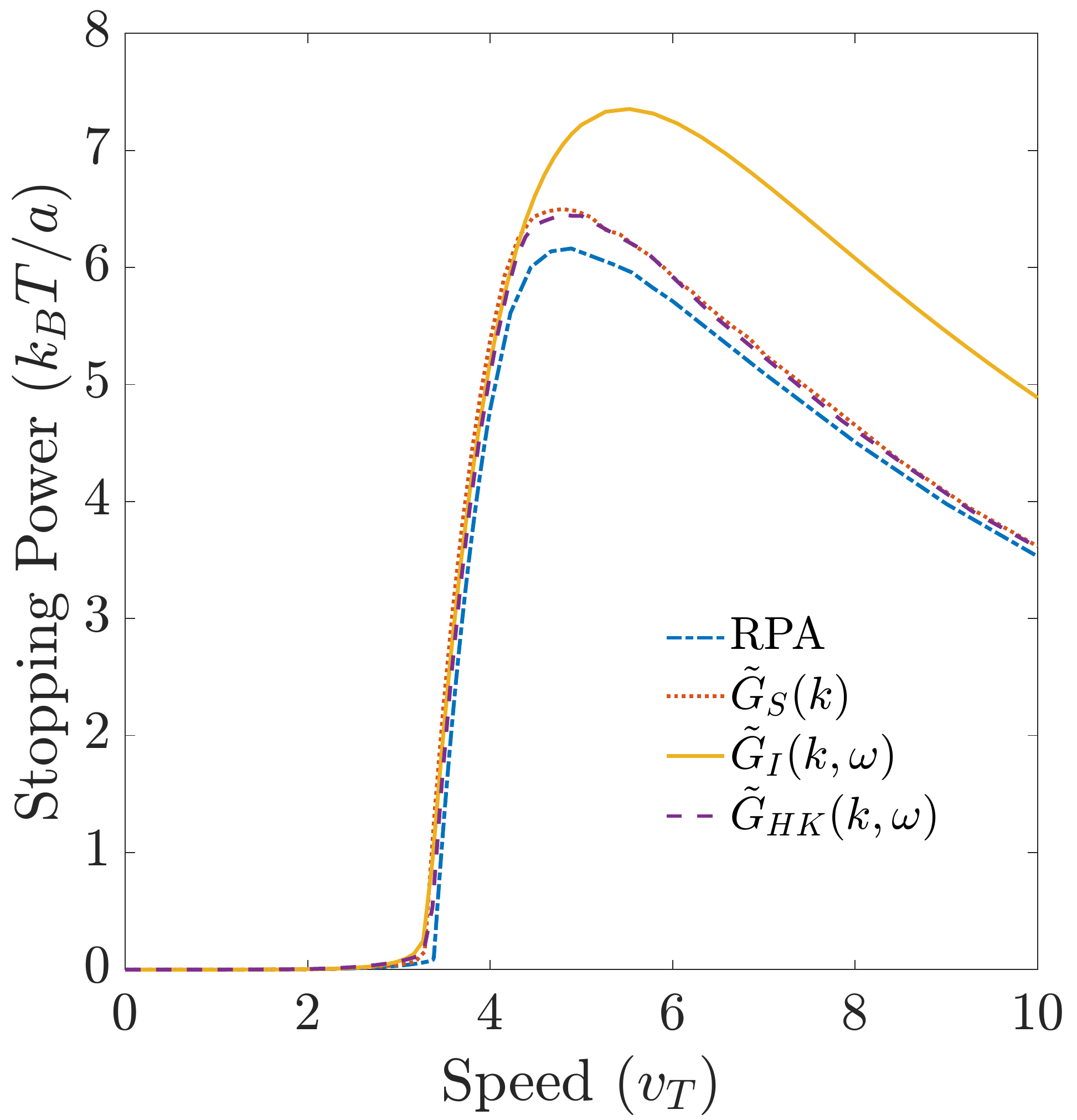}
\caption{Stopping power for $\massrat=1000$ at $\Gamma=10$ computed from Eq.~(\ref{eq:dEdx_dielectric}) using each of the four model dielectric functions: RPA from Eq.~(\ref{eq:weakeps}) (blue dashed line), and LFC models from Eq.~(\ref{eq:LFC_dielectric}) with the static approximation from Eqs.~(\ref{eq:LFC}), Ichimaru dynamic LFC from Eq.~(\ref{eq:Ich_DLFC}) (solid yellow line) and Hong-Kim dynamic LFC from Eq.~(\ref{eq:HK_DLFC}) (dashed purple line).}
\label{fig:DLFC_Compare}
\end{figure}

Figure~\ref{fig:DLFC_Compare} shows a comparison of the stopping power calculated from Eq.~(\ref{eq:dEdx_dielectric}) at conditions $\Gamma = 10$ and $\massrat = 1000$ for each of the four models for the dielectric function. 
The figure shows that the prediction of each model is essentially identical for speeds below the Bragg peak. 
For speeds above the Bragg peak, the dynamic LFC model from Eq.~(\ref{eq:Ich_DLFC}) predicts a stopping power that is approximately 20\% larger than the other models. 
Since it is unknown which of the dynamic LFC models is more accurate, and the predictions are quite similar anyway, only the static LFC model is compared with the MD data.  

\acknowledgments
The authors are grateful for helpful discussions with Dr. N. R. Shaffer and Dr. G. H. Bernstein.
 
This material is based upon work supported by the U.S. Department of Energy, Office of Science, Office of Fusion Energy Sciences under Award Number DE-SC0016159 and by the National Science Foundation under Grant No.~PHY-1453736. 
It used the Extreme Science and Engineering Discovery Environment (XSEDE), which is supported by NSF Grant No. ACI-1053575, under Project Award No. PHYS-150018. 
The work of JD  was supported by the US Department of Energy through the Los Alamos National Laboratory and by LDRD Grant No. 20170073DR. Los Alamos National Laboratory is operated by Triad National Security, LLC, for the National Nuclear Security Administration of U.S. Department of Energy (Contract No. 89233218CNA000001).

\bibliography{refs.bib}

\end{document}